\UseRawInputEncoding
\pdfoutput=1
\documentclass[11pt]{article}
\usepackage{jheppub}
\usepackage{amsthm}
\usepackage{amsmath}
\usepackage{amssymb}
\usepackage{braket}
\usepackage{multirow}
\usepackage{graphicx}
\usepackage{hyperref}
\usepackage{mathrsfs}
\usepackage{bm}
\usepackage{subcaption}
\usepackage{empheq}
\usepackage{appendix}
\usepackage{adjustbox}
\usepackage{natbib}
\usepackage[dvipsnames]{xcolor}
\usepackage{float}
\usepackage{slashed}
\usepackage{indentfirst}
\graphicspath{{Figs/}}
\usepackage{cleveref}
\usepackage{hyperref, pdfsync, epstopdf, enumerate}
\DeclareMathOperator{\vol}{vol}
\DeclareMathOperator{\arctanh}{ArcTanh}
\DeclareMathOperator{\ArcTan}{ArcTan}
\DeclareMathOperator{\sign}{sgn}
\hypersetup{colorlinks,bookmarksopen,bookmarksnumbered,
citecolor={Red},
linkcolor={Red},pdfstartview=FitH,
urlcolor={Red}}

\begin{document}
\title{Non-integrability in AdS$\bm{_3}$ vacua}
\author[a]{Konstantinos S.~Rigatos,}
\affiliation[a]{School  of  Physics  \&  Astronomy  and  STAG  Research  Centre,  University  of  
Southampton, Highfield,\\ Southampton  SO$17$ $1$BJ, UK.}
\emailAdd{k.c.rigatos@soton.ac.uk}
\dedicated{Dedicated to the memory of Michalis Kaltezas. He was only 15. Never forget, never forgive. $\bm{1312}$.}
\abstract{We ask the question of classical integrability for certain (classes of) supergravity vacua that contain an AdS$_3$ factor arising in massive IIA and IIB theories and realizing various and different amounts of supersymmetry. Our approach is based on a well-established method of analytic non-integrability for Hamiltonian systems. To detect a non-integrable sector we consider a non-trivially wrapped string soliton and study its fluctuations. We answer in the negative for each and every one of the supergravity solutions. That is, of course, modulo very specific limits where the metrics reduce to the AdS$_3 \times$ S$^3 \times \tilde{S}^3 \times$ S$^1$ and AdS$_3 \times$ S$^3 \times$ T$^4$ solutions which are known to be integrable.}
\maketitle                                                                          
\setcounter{page}{1}\setcounter{footnote}{0}
\newpage
\renewcommand*{\thefootnote}{\arabic{footnote}}
\section{Prolegomena}
	\subsection{Motivation}
The AdS/CFT duality \cite{Maldacena:1997re,Witten:1998qj,Gubser:1998bc} constitutes a bona fide realization of the holographic principle. The correspondence postulates that specific string theories with an AdS factor in their geometry are equivalent descriptions of field theories residing on the boundary of the space. While the original proposal related type IIB string theory in $AdS_5 \times S^5$ to the four-dimensional $\mathcal{N}=4$ super Yang-Mills theory, it was generalized shortly after its discovery to more general gauge/gravity paradigms in diverse dimensions \cite{Itzhaki:1998dd}. Many years after its discovery and with all the effort that has been put towards the more precise understanding of its aspects, the duality has proved iteself to be a powerful tool that allows us to probe the dynamics both in the field theory side and in the gravity side.  

This has enacted a large and sundry avenue of explorations and questions. A salient branch of the AdS/CFT has been the meticulous study of superconformal field theories in four dimensions in conjunction with their higher and lower dimensional counterparts. Note that superconformal field theories can exist if the dimensionality of the spacetime is no bigger than six \cite{Nahm:1977tg}. A lot of progress has been made towards the complete classification of supergravity backgrounds in Type II and M-theory that contain an AdS$_{d+1}$ submanifold; see for instance \cite{Gauntlett:2004zh,Gutowski:2014ova} and references therein. These vacua are conjectured to be the holographic duals to $d$-dimensional superconformal field theories with disparate supercharges.  

Four dimensional $\mathcal{N}=2$ theories were minutely scrutinized from the field theory viewpoint \cite{Gaiotto:2009we} with their holographic gravity descriptions obtained in \cite{Gaiotto:2009gz} and further elaborations on these theories were also pursued; see as representatives \cite{ReidEdwards:2010qs,Aharony:2012tz,Bah:2015fwa,Bah:2019jts,Nunez:2018qcj,Nunez:2019gbg} amongst other works. Five dimensional superconformal field theories with various amounts of supersymmetry are investigated in \cite{Lozano:2012au,DHoker:2016ujz,DHoker:2016ysh,DHoker:2017mds,Gutperle:2018vdd,Fluder:2018chf,Bergman:2018hin,Lozano:2018pcp}, and an infinite class of six-dimensional $\mathcal{N}=(1,0)$ theories was likewise inspected in the works \cite{Brunner:1997gf,Hanany:1997gh,Gaiotto:2014lca,Cremonesi:2015bld,Apruzzi:2013yva,Apruzzi:2014qva,Apruzzi:2015wna,Apruzzi:2017nck,Passias:2015gya,Bobev:2016phc,Macpherson:2016xwk,Filippas:2019puw}. Superconformal quantum mechanics are also being under consideration in the literature for quite some time. Representative examples of the said studies include \cite{Lozano:2020txg,Cvetic:2000cj,Gauntlett:2006ns,DHoker:2007mci,Gupta:2008ki,Kim:2013xza,Corbino:2017tfl,Dibitetto:2018gbk,Dibitetto:2018gtk,Hong:2019wyi,Dibitetto:2019nyz,Lust:2020npd}.

The specific case of the AdS$_3$/CFT$_2$ is particularly alluring. The reason is that two dimensional (super)conformal field theories and three dimensional AdS spacetimes make their appearence in various places in theoretical physics; string theory, condensed matter physics, black holes to name a few. The power of the (super)conformal algebra in two dimensions provides an excellent theoretical playground where we are allowed to examine and test a number of different ideas explicitly. All of the above are motivating factors for finding and classifying AdS$_3$ supergravity backgrounds in addition to studying their boundary filed theoretic descriptions. There is a vast and rich literature related to these aspects and since we cannot do justice to it, here we mention some exemplaries \cite{Witten:1997yu,Seiberg:1999xz,Kutasov:1998zh,Hohenegger:2008du,Maldacena:1997de,Martelli:2003ki,Castro:2008ne,Couzens:2019wls,Kim:2015gha,Gadde:2015tra,Lawrie:2016axq,Couzens:2017way,Lozano:2015bra,Kelekci:2016uqv,Macpherson:2018mif,Dibitetto:2018ftj,Lozano:2019emq,Lozano:2019zvg,Filippas:2020qku}. 

In cognate with the above, integrability possesses a protagonistic role towards the more comprehensive understanding of field theories inasmuch as it models the physical aspects of the systems that exhibit the homonymous property. Its appearence unveils a plethora of structures of conserved quantities. Owing to this, the theory is solvable for any value of the gauge coupling. This ties in neatly with the precedent discussion as we can link the superstring worldsheet description to a superconformal field theory without gravity using holography, and as a consequence the integrability of the boundary gauge theory naturally translates to a question in the gravity side of the duality for the integrability of the string.

The most prominent computations of the best understood AdS/CFT example between the $AdS_5 \times S^5$ supergravity solution and the $\mathcal{N}=4$ super Yang-Mills rely on the full integrability of the system \cite{Beisert:2010jr}. The classical string integrability in the aforementioned example is exhibited as a flat condition on the Lax connection through the Lagrangian  equations of motion \cite{Bena:2003wd}. The same statement is shown to also hold true for strings that propagate in the Lunin-Maldacena background \cite{Lunin:2005jy}, which is the dual description to the marginal Leigh-Strassler deformation, with a $\beta$-parameter that is real and that preserves $\mathcal{N}=1$ supersymmetry \cite{Frolov:2005dj}. However, the deformation parameter $\beta$ is allowed to take complex values as well and in that more general setup, integrability is lost \cite{Frolov:2005ty,Berenstein:2004ys,Giataganas:2013dha}.

In spite of the fact that integrable field theories possess a number of delectable features, it is highly cumbersome to deduce that a specific theory is integrable. The reason is that integrability is related to the existence of the Lax connection, and to this day there is no algorithmic and systematic way to construct it. To be more precise, there is absolutely no reason to determine whether or not such a connection exists a priori. As a result, integrable theories are obtained predominantly as deformations that preserve some structures of theories that are known to be integrable \cite{Sfetsos:2013wia,Delduc:2014kha,Borsato:2016pas}.

Due to the objective hindrances in spotting integrable structures, combined with the fact that integrability has to manifest itself universally in a given theory, reverse engineering the logic of integrability by searching for non integrable dynamics arises dialectically. The complete analysis comprises of the study of the non-linear partial differential equations that arise from the string $\sigma$-model. In practice, a more resilient approach is to study certain wrapped string embeddings and then analyse the equations of motion that result from it. Since integrability is a propery that has to be manifested universally in a theory, a single counter-example is sufficient to declare the full theory as being non-integrable. This method has been dubbed analytic non-integrability. 

A substantially different approach has been used as well in order to obtain appropriate conditions of non-integrability is the factorization of the S-matrix on the worldsheet of the theory \cite{Wulff:2017lxh,Wulff:2017vhv,Wulff:2017hzy,Wulff:2019tzh}. The method of analytic non-integrability was originally developed in \cite{Zayas:2010fs} and the recent work of \cite{Giataganas:2019xdj} provides a bridge between these two distinct approaches of non-integrability.  

In the absence of a complete and general classification of two-dimensional integrable $\sigma$-models, proving the non-integrability of certain models is also an important asset to contemporary research. It provides an essential step towards carving out the phase space on integrable and non-integrable models. On top of that, searching for and classifying integrable subsectors within non-integrable theories remains a principal and key question in the mapping of the transition to non-integrability from integrable systems. 

The procedure of analytic non-integrability was originally proposed in \cite{Zayas:2010fs}. That method has been used a number of times in the past \cite{Basu:2011di,Basu:2011fw,Rigatos:2020hlq,Stepanchuk:2012xi,Nunez:2018qcj,Nunez:2018ags,Filippas:2019ihy,Filippas:2019bht,Giataganas:2017guj,Chervonyi:2013eja,Roychowdhury:2017vdo,Giataganas:2014hma,Roychowdhury:2019olt,Banerjee:2018ifm,Filippas:2019puw} very succesfully when examining different supergravity theories with various amounts of symmetry. 

It is worthwhile stressing that even if a specific background is determined as being generally non-integrable, this does not preclude the possible existence of integrable subsectors in the given theory. A very neat illustrative example of this  is provided by the complex $\beta$-deformation. We have already seen that this theory is non-integrable. Be that as it may, the subsector that is made out of two holomorphic and an antiholomorphic scalar field is known to be one-loop integrable \cite{Mansson:2007sh}. Not only that, but also fast spinning strings in the same subsector with a purely imaginary deformation parameter are also integrable \cite{Puletti:2011hx}.  

Furthermore, while string theory in five-dimensional Sasaki-Einstein manifolds has been shown to be non-integrable generally \cite{Basu:2011di,Basu:2011fw,Rigatos:2020hlq} since the string exhibits non-integrable dynamics and chaotic motion in these solutions, there are recent studies on integrable deformations thereof \cite{Arutyunov:2020sdo,Rado:2020yhf}. 

In this work we contribute to the classification of non-integrable supegravity solutions. We are considering two classes of supergravity backgrounds in the massive IIA and IIB theories with large and small $\mathcal{N}=(4,0)$ superconformal symmetry respectively \cite{Macpherson:2018mif} and additionally the unique local solution that realizes the exceptional $F(4)$ maximal superlagebra \cite{Dibitetto:2018ftj}. In the first two cases the backgrounds are parameterized in terms of constants, which naturally begs the question of a potential fine tuning of these constants that would lead to an integrable theory. By enforcing the full power of the analytic Kovacic algorithm \cite{KOVACIC19863} as we explain in the following sections, we exclude such a possibility and thus we determine that all of our cases are completely non-integrable. 
	\subsection{A brief synopsis of the method}
The method we adopt here is comprised out of the following steps: to begin with, we write an ansatz for a string embedding that extends in some dimensions and it wraps non-trivially some of the cyclic coordinates in a given background. An indication for the consistency of our choice is to derive the equations of motion for the said string configuration that follow from the analysis of the bosonic $\sigma$-model lagrangian and the associated wordlsheet energy-momentum tensor and subsequently show that the Virasoro conditions are satisfied on-shell; when we enforce the E\"uler-Lagrange equations. 

The said string soliton contains $\mathcal{X}$ degrees of freedom. We wish to obtain simple solutions for the $(\mathcal{X}-1)$ equations of motion. These simple solutions define the so-called invariant plane of solutions. Having solved all the equations of motion but one, we focus on this last one. We allow small fluctuations in that dimension and on the invariant plane of solutions. In other words, we freeze all the dimensions on the values that solve their equations of motion and only then consider fluctuations in the coordinate under examination.

In such a way, we arrive at second-order, ordinary differential equation that assumes the schematic form 
\begin{equation}
\mathcal{A}_1 ~ f'' + \mathcal{A}_2 ~ f' + \mathcal{A}_3 ~ f = 0\, .
\end{equation}

The above is called the Normal Variational Equation (NVE). If the coefficients of the NVE, $\mathcal{A}_i$ for $=1,2,3$, are not rational functions we need to perform change of coordinates and/or perhaps other algebraic manipulations to bring them in an appropriate form.

In this kind of Hamiltonian dynamical systems, in order to make a statement for the analytic non-integrability of the structure of the aferementioned systems, one has to enforce differential Galois theory. The Galois theory on differential equations was understood how to be, equivalently, turned into an algebraic statement in \cite{KOVACIC19863}. In that work, Kovacic derived an explicit algorithm that produces the Liouvillian solutions -if they exist- of a second-order, ordinary,  linear differential equation; the kind of equations that describes these systems.   

The relation of the Kovacic algorithm to differential Galois theory and the full analytic Kovacic algorithm that produces the Liouvillian solutions have been presented in the literature a number of times before, and so we refrain from presenting these basic formulae here once more. In our opinion, the most prominent and thorough presentation can be found in Appendix A of \cite{Filippas:2019ihy}.

Since we have presented the nuts and bolts of the method that we adopt in this work, we feel that it is appropriate to make a comment regarding its validity. 

This is a method that we use such that we prove that a certain sub-sector of theory that we are examining is non-integrable. Differential Galois theory through the Kovacic algorithm is a statement for the Liouville (non)-integrability of an NVE. This means that if a certain NVE is non-integrable, then it is perfectly valid to declare that the string theory -and by menas of the AdS/CFT the boundary field theory description- to be non-integrable. However, if we end up having Liouville integrable solutions for all the NVEs, \textit{it does not necessarily} mean that the theory is itself integrable. It does not even mean that the sub-sector under examination is clasically integrable. As we will see, the string motion is described by coupled differential equations, and decoupling them to arrive at the NVEs is just a mathematical trick. So, one ought to be cautious before making the statement of integrability. It is, of course, a very encouraging indication to obtain Liouvillian solutions through this method, but much more is needed to prove the integrability of a certain vacuum.

On the other hand, if a given theory or for that matter a certain sub-sector of a supergravity description has been shown to be integrable, then this fact \textit{should} be reflected upon the NVEs. The way that this occurs is that the NVEs will assuredly have simple solutions in terms of quadratures.  
	\subsection{Comments on the field theoretical realization} \label{sec: field_theory_comments}
At this point we feel that it is appropriate to pause and discuss a subtle issue that is not usually emphasized in this kind of analysis. A happy exception to the said rule can be found in \cite{Filippas:2019bht} in the context of non-integrability of the $\Omega$-deformed $\mathcal{N}=4$ super Yang-Mills. 

In the kind of analysis of the dynamics for a given string soliton that we undertake here, one usually obtains an  equation of motion for the time coordinate of the target spacetime of appropriate form, such that it gives the string's energy as its first integral. This is a point of the utmost importance in order to be albe to argue using holography that a specific string configuration is a state that can be thought of as the gravity dual to a gauge invariant operator. Being able to argue in the way we just sketched constitutes a necessity if we desire to share the statement of (non)-integrability from the string side to the boundary gauge theory. Since we always incorporate the worldsheet time coodinate in the ten-dimensional time coordinate via $t=t(\tau)$ and this enters the dynamics through the equations of motion derived from the $\sigma$-model, everything is apt and our string configuration should have a well-defined holographic realization.

In simple words, we should care about the consistency of our string soliton for a given (class of) supergravity background(s). Evidently, in the foregoing argument the precise knowledge of the particular form of the operator is not a necessary requirement. The same holds true for the boundary superconformal field theory as well. We do not need to know precisely the holographic dictionary to share the statement of non-integrability so long as our string embedding is consistent. 

All the vacua that we consider in this work are gravity descriptions of as of yet undetermined boundary field theoretic descriptions. However, as we shall see, using the aforementioned discussion -the validity of the wrapped string we examine- we will be able to share the property of Liouville (non)-integrability with the unspecified superconformal field theory. 

Related to the field theory interpretation, but in a different way, we also feel that another explanatory comment is in order. This time it has to do with the form of the boundary operators.

We will be working with extended strings that wrap non-trivially some cyclic coordinates of the geometry. Usually, the holographic field theory opeators that correspond to these string states are long, unprotected operators with large quantum numbers, i.e large energy and/or angular momentum. In the case of the Klebanov-Witten model this has been more thoroughly elaborated in \cite{Basu:2011di}, since we have a better understanding and control over the boundary superconformal field theory. 

In a nutshell, the moral of the argument regarding the heuristic form of the operators in the boundary field theory that are dual to the extended wrapped strings whose dynamics we consider is that there are BMN-like operators with impurities inserted such that they carry appropriate quantum numbers to be associated with the angular momenta along the cyclic coordinates around which we wrap the string soliton.

Of course, a precise and more elaborate discussion of such strings and the matching the BMN-like operators is given in \cite{Basu:2011di} for string theory in the $AdS_5 \times T^{1,1,}$. Note that in order to make so precise statements requires the knowledge of the relation between the field content and the classical variables. For the Klebanov-Witten model (string theory in $AdS_5 \times T^{1,1}$) this has been discussed in \cite{PandoZayas:2002dso}. 

In the limit where we take the wrapping of the sting to be trivial, we end up examining the low-energy description of the string which appears as a point particle that is moving along geodesic paths.  
	\subsection{The structure of this work}
The structure of this work is hopefully such that it makes navigations easy. \Cref{sec: string dynamics} contains the main formulae that we will need to use throughout this work. More specifically, \cref{sec: no_nsns_sector} contains all the necessary relations in terms of general and unspecified warp metric factors that can be applied to all of the supergravity vacua considered explicitly in this work. Specifically this analysis applies to \cref{sec: largesusy,sec: IIB_vacua,sec: exceptional}. In \cref{sec: ads3s3s3s1} we explain and show how the integrability of a given supergravity solution is manifested in the method of analytic non-integrability that we utilize here. We procced to analyze the first class of supergravity descriptions that realize large $\mathcal{N}=(4,0)$ in massive IIA. This analysis can be found in \cref{sec: largesusy}. We, then, move on to IIB solutions with small $\mathcal{N}=(4,0)$ in \cref{sec: IIB_vacua}. \Cref{sec: exceptional} is devoted to the discussion of the AdS$_3$ supergravity that realizes the maximal, $\mathcal{N}=8$, exceptional $F(4)$ superalgebra. In \cref{sec: final} we summarize our findings. Finally, we supplement the material of the main body with \cref{app: analytic_kovacic}, where go through the steps of the Kovacic algorithm explicitly for the parameterized differential equations arising from the two classes of backgrounds in \cref{sec: largesusy,sec: IIB_vacua}.
\section{String dynamics} \label{sec: string dynamics}
The Polyakov action which depicts the bosonic string dynamics is given by\footnote{in the case of a non-trivial NS-NS sector one should add to the action the term $\epsilon^{\alpha \beta} ~ B_{MN}~\partial_{\alpha} X^{M} \partial_{\beta} X^{N}$}
\begin{align} \label{eq: polyakov_action}
\begin{aligned}
S = \frac{1}{4 \pi \alpha'} \int d^2 \sigma ~    h^{\alpha \beta} ~ G_{MN} ~ \partial_{\alpha} X^{M} \partial_{\beta} X^{N}\, ,
\end{aligned}
\end{align}
in the conformal gauge and the equations of motion for the string coordinates, $X^{M}(\tau, \sigma)$, that follow from the above are constrained by the Virasoro conditions. The worldsheet energy-momentum tensor is expressed via
\begin{equation} \label{eq: energy_momentum_tensor}
T_{\alpha \beta} = \frac{1}{\alpha^{\prime}} \left( g_{MN} ~ \partial_{\alpha} X^M \partial_{\beta} X^N - \frac{1}{2} g_{MN} ~ \eta_{\alpha \beta} ~ \eta^{\gamma \delta} ~ \partial_{\gamma} X^M \partial_{\delta} X^N \right)\, ,
\end{equation}
and using the above definition for the worldsheet energy-momentum tensor, we can write the Virasoro conditions explicitly in the following way
\begin{equation} \label{eq: virasoro_constraints}
\begin{split}
T_{\tau \sigma} &= T_{\sigma \tau} = G_{MN} \dot{X}^{M} \acute{X}^{N} = 0, \\
2T_{\tau \tau} &= 2 T_{\sigma \sigma} =G_{MN} \left( \dot{X}^{M} \dot{X}^{N} + \acute{X}^{M} \acute{X}^{N} \right) = 0\, ,
\end{split}
\end{equation}
where we have used the abbreviations $\dot{X} \equiv \partial_{\tau} X$ and $ \acute{X} \equiv  \partial_{\sigma} X$. 

We keep in mind that we want to enforce differential Galois theory on the equations of motion for our string embeddings. This means that we need to end up having an ordinary differential equation resulting from our studies of the string's dynamics. Simply put, we need to choose for the string coordinates to be either of the form $X^M = X^M(\tau)$ or $X^M=X^M(\sigma)$ with $\tau$ and $\sigma$ being the worldsheet time and spacelike coordinates respectively. We want to bring the dynamics of the string soliton to test in order to examine the (non)-integrability and this is why we will allow the string to wrap non-trivially cyclic coordinates. Note that it is precisely this wrapping that reinforces the stringy character of our embedding. Without this wrapping, we would have the point-like limit of the string, which corresponds to the low-energy supegravity description of a specific background. 

We will find it very useful to pass to the Hamiltonian formulation of systems under consideration and doing so also provides some great physical intuition on the systems under examination. To do so, we employ the usual relations for the canonical conjugate momenta which are expressed via 
\begin{equation} \label{eq: canonical_conjugate_momenta_gen}
p_M = \frac{\partial \mathcal{L}}{\partial \dot{X}_M} \, ,
\end{equation} 
and using the above the Hamiltonian can be evaluated using 
\begin{equation} \label{eq: hamiltonian_general}
\mathcal{H} = \sum_A \dot{q}_A ~ p_A - \mathcal{L} \, ,
\end{equation}
where the above we use $q^A$ to denote the generalized coordinates of our systems.

In the next section, see \cref{sec: no_nsns_sector}, we provide an analysis in general terms with unspecified warp factors. While we are analyzing the $S^3 \times S^3$ fibration that was utilized by the authors of \cite{Macpherson:2018mif} in (massive) IIA and IIB theories, the same relations can also be used in the vacuum with exceptional supersymmetry that are of the schematic form $AdS_3 \times \mathcal{I} \times S^6$, since we can consider three of the angles to have a non-trivial profile and the string to be localized at certain points along the other three. 
	\subsection{The $S^3 \times \tilde{S}^3$ fibration with no NS-flux} \label{sec: no_nsns_sector}
The backgrounds that we consider here have only an RR sector and a dilaton and are described schematically by an invariant line element of the form\footnote{up to some unimportant constants relating the warp factors.}
\begin{equation} \label{eq: general_geometry_1}
ds^2 = h_0 ~ ds^2_{AdS_3} + \frac{1}{h_0} ~ dr^2 + h_1 ~ ds^2_{S^3} + h_2 ~ ds^2_{\tilde{S}^3}\, ,
\end{equation}
where in the above the various unspecified warp factors are functions of the $r$-coordinate, $h_i = h_i(r)$ for $i=0,\cdots,2$, and we have not written the dilaton and the RR fields as they are not of relevance for our purposes. We choose to express the $AdS_3$ spacetime in global coordinates. For unit radii for the three-dimensional $AdS$ spacetime and the two different three-spheres, the geometries are given by
\begin{equation} \label{eq: general_geometry_2}
\begin{aligned}
ds^2_{AdS_3} 		&= - \cosh^2 \rho dt^2 + d \rho^2 + \sinh^2 \rho dw^2 \, , \\
ds^2_{S^3} 			&= d \theta^2_1  + \sin^2 \theta_1 (d \theta^2_2 + \sin^2 \theta_2 d \theta^2_3)\, ,\\
ds^2_{\tilde{S}^3}	&= d \phi^2_1  + \sin^2 \phi_1 (d \phi^2_2 + \sin^2 \phi_2 d \phi^2_3)\, ,
\end{aligned}
\end{equation}
where the coordinates take values within the following ranges: $0 \leq \{w, \theta_1, \phi_1 \} \leq \pi/2$ as well as $0 \leq \{\theta_2, \phi_2, \theta_3, \phi_3 \} \leq 2\pi$.

We consider our string embedding to be desribed by the following ansatz 
\begin{equation} \label{eq: string_ansatz_1}
\begin{aligned}
t 		&= t(\tau)\, , 			&&& r	&= r(\tau)\, ,		&&& \theta_1 &= \theta_1(\tau)\, , \\
\rho 	&= \rho(\tau)\, ,		&&& {}	&{} 				&&& \theta_2 &= \alpha_2 \sigma\, ,\\
w 		&= \alpha_1 \sigma\, ,	&&& {} 	&{}					&&& \theta_3 &= \frac{\pi}{2}\, .
\end{aligned}
\end{equation}

It is clear from the above that the string configuration is wrapped around the $w$-coordinate $\alpha_1$ times and around the $\theta_2$ dimension $\alpha_2$ times. We allow the string to pulsate along the remaining $AdS_3$, $S^3$ and $r$ coordinates while we consider it to be placed at certain constant points in the second three-sphere, $\tilde{S}^3$, i.e $\phi_1=\phi_2=\phi_3=\pi/2$. The fact that $\alpha_1$ and $\alpha_2$ are non-trivial ensures the stringy character of the embedding. In the special case $\alpha_1 = \alpha_2 = 0$ the string degenerates to a point-particle. 

Instead of working with the action of the non-linear $\sigma$-model, \cref{eq: polyakov_action}, it is easier to consider its associated Lagrangian density. When evaluated for the particular choice of a string embedding that we made above, \cref{eq: string_ansatz_1}, it is given by 
\begin{equation} \label{eq: lagrangian_gen_1}
\mathcal{L} = h_0 ~ (\cosh^2 \rho ~ \dot{t}^2 - \dot{\rho}^2 + \sinh^2 \rho ~ \alpha^2_1) - \frac{1}{h_0} ~ \dot{r}^2 - h_1 (\dot{\theta_1}^2 - \sin^2 \theta_1 \alpha^2_2)\, .  
\end{equation}
The equations of motion that follow from the above Lagrangian density for the various coordinates are given by
\begin{equation} \label{eq: eom_gen_1}
\begin{aligned}
\dot{t} 		&= \frac{E}{h_0 ~ \cosh^2 \rho} \, , \\
\ddot{\rho}		&=-\frac{\partial_r h_0}{h_0} ~ \dot{r} ~ \dot{\rho} - \cosh \rho ~ \sinh \rho \left(\alpha^2_1 + \frac{E^2}{h^2_0 ~ \cosh^4 \rho} \right) \, ,\\
\ddot{r}		&=- \frac{\partial_r h_0}{2 h_0} \left( \dot{r}^2 - \frac{E^2}{\cosh^2 \rho} \right) \\ 
&+ \frac{1}{2} h_0 \left(\vphantom{\frac{1}{2}} (\partial_r h_0) (\dot{\rho}^2 - \sinh^2 \rho ~ \alpha^2_1) + (\partial_r h_1) (\dot{\theta_1}^2 - \sin^2 \theta_1 ~ \alpha^2_2) \right) \, ,\\
\ddot{\theta_1}	&=-\frac{\partial_r h_1}{h_1} ~ \dot{r} ~ \dot{\theta_1}- \cos \theta_1 ~ \sin \theta_1 ~ \alpha^2_2 \, ,
\end{aligned}
\end{equation}
where in the above $E$ is just a constant and we have replaced the expression for $\dot{t}$ in the remaining equations of motion. 

The equations of motion derived above are constrained by the Virasoro conditions, the worldsheet equations of motion, which in this case read 
\begin{equation} \label{eq: energy_momentum_gen_1}
\begin{aligned}
2 T_{\tau \tau} = 2 T_{\sigma \sigma} &= h_0 ~ (-\cosh^2 \rho ~ \dot{t}^2 + \dot{\rho}^2 + \sinh^2 \rho ~ \alpha^2_1) + \frac{1}{h_0} ~ \dot{r}^2 + h_1 ~ (\dot{\theta_1}^2 + \sin^2 \theta_1 \alpha^2_2)\, , \\
T_{\tau \sigma} = T_{\sigma \tau} &= 0 \, .
\end{aligned}
\end{equation}

The above constraints have to hold true irregardless of the Lagrangian equations of motion. Truely, the worldsheet energy-momentum tensor given by \cref{eq: energy_momentum_gen_1} is conserved when evaluated on-shell -in other words when evaluated on the equations of motion given by \cref{eq: eom_gen_1}. That is, we have $\nabla^{\alpha}T_{\alpha \beta}=0$, since $\partial^{\tau}T_{\tau \tau}=\partial^{\sigma}T_{\sigma \sigma}=0$ when we enforce the equations of motion \cref{eq: eom_gen_1} on these expressions. 

It is worthwhile  pointing out that this agreement between the worldsheet equations of motion (the Virasoro conditions) and the equations of motion for the target spacetime coordinates is indicative for the consistency of our string soliton, \cref{eq: string_ansatz_1}.

We now wish to turn to the Hamiltonian formulation of the systems considered here. To begin with, we compute the canonical conjugate momenta using \cref{eq: canonical_conjugate_momenta_gen} which read
\begin{equation} 
\begin{aligned}
p_t &= 2 ~ h_0 ~ \cosh^2 \rho ~ \dot{t} \, ,	&&&	p_{\rho} &= -2 ~ h_0 ~ \dot{\rho} \, ,		&&& p_r &= -\frac{2}{h_0}~\dot{r} \, ,		&&& p_{\theta_1} &= 2~h_1~\dot{\theta_1} \, ,
\end{aligned}
\end{equation}
and we can use the above to derive the relevant Hamiltonian density. It is explicitly equal to 
\begin{equation}
\mathcal{H} = \frac{p^2_t}{4~h_0~\cosh^2\rho} - \frac{p^2_{\rho}}{4~h_0} - \frac{h_0}{4}~p^2_r - \frac{p^2_{\theta_1}}{4~h_1} - h_0 ~ \sinh^2 \rho ~ \alpha^2_1 - h_1~\sin^2\theta_1~\alpha^2_2 \, .
\end{equation}
In this Hamiltonian formulation the Virasoro conditions are equivalent to the statement $\mathcal{H}=0$. The equations that follow from the aforementioned Hamiltonian density are, of course, identical to E\"uler-Lagrange equations of motion \cref{eq: eom_gen_1}. Having said that, it is quite straightforward to give a neat and illustrative classical mechanics explanation of the system under consideration here. The string dynamics has effectively reduced to that of a particle moving in the presence of a non-trivial potential. The non-trivial wrapping of string (the winding around the cyclic coordinates) is responsible for generating the said potential. The effective mass can read off from the kinetic terms which are due to the geometry under examination.

The dynamics of our string soliton are delineated by an involved system of differential equations. The reason is that the equations are inherently coupled. Working and trying to solve such a system is a formidable task in general. Be that as it may, there is an elegant way to bypass this complication and facilitate our needs for the forthcoming analysis. To that end, we want to find a simple solution to the equations of motion that the string satisfies, see \cref{eq: eom_gen_1}, and allow the coordinates to fluctuate around them. Such a fluctuation around the set of simple solutions is called the NVE for a given coordinate. At this point we want to remind the reader and stress that the Virasoro condition, \cref{eq: energy_momentum_gen_1}, is a primary constraint. Having said that, the simple solution to the equations of motion should satisfy these conditions as well. 

By inspecting the string's dynamics it is quite easy to see that the following simple solutions\footnote{it should also be quite obvious that this plane of solutions is not unique and we chose it for convenience in the manipulations.}
\begin{equation} \label{eq: gen_invariant_plane_1}
\rho = \dot{\rho} = \ddot{\rho} = \theta_1 = \dot{\theta_1} = \ddot{\theta_1} = 0\, ,
\end{equation}
automatically satisfy the equations of motion for the $\rho$ and $\theta_1$ coordinates. On this plane of solutions, the equation of motion that the $r$-dimension satisfies gets simplified to
\begin{equation}
\ddot{r} = \frac{\partial_r h_0}{2~h_0}~(\dot{r}^2-E^2) \, ,
\end{equation}
which can be obviously solved by the simple expression
\begin{equation} \label{eq: general_simple_sltn_r_1}
r = \bar{r} = E ~ \tau\, .
\end{equation}

It is a simple matter of a straightforward computation to evaluate the Virasoro condition \cref{eq: energy_momentum_gen_1} on the solutions we just found described by \cref{eq: gen_invariant_plane_1} that satisfy the Lagrangian equations of motion. Doing so yields
\begin{equation}
2 T_{\tau \tau} = 2 T_{\sigma \sigma} = \frac{1}{h_0} (\dot{r}^2-E^2) = 0\, ,
\end{equation} 
which is obviously solved by \cref{eq: general_simple_sltn_r_1}.

Hence we have shown that all the pieces thus far come as advertised and the simple solutions that we managed to find and subsequently consider fluctuations around them describe the same dynamics whether we examine the worldsheet equations of motion or the equations of motion from the $\sigma$ model's Lagrangian.

We have set up the scene to derive the NVEs which are the quantities of main interest. We begin by examining fluctuations of the $\rho$-coordinate around its simple solution, $\rho=0+\varepsilon~\varrho$ with $\varepsilon \rightarrow 0$ and obtain 
\begin{equation} \label{eq: NVE_general_1a}
\mathcal{W}_0 ~ \ddot{\varrho} + \mathcal{W}_1 ~ \dot{\varrho} + \mathcal{W}_2 ~\varrho = 0\, ,
\end{equation}
where the prefactors in the above expression are given by:
\begin{equation} \label{eq: NVE_general_1a_factors}
\begin{aligned}
\mathcal{W}_0 &= \left.\vphantom{\frac{1}{2}} h_0 \right\vert_{r=\bar{r}}, &&& \mathcal{W}_1 &=  E \left.\vphantom{\frac{1}{2}} \left( \partial_r h_0 \right) \right\vert_{r=\bar{r}}, &&& \mathcal{W}_2 &= \left.\vphantom{\frac{1}{2}} \left( \alpha^2_1 + \frac{E^4}{h^2_0} \right) \right\vert_{r=\bar{r}}\, .
\end{aligned}
\end{equation}

We can work in a similar manner to obtain the NVE pertaining to the $\theta_1$-dimension. We expand around its simple $\theta_1 = 0 + \varepsilon~\vartheta$ and working in the $\varepsilon \rightarrow 0$ limit. Working to leading order in the small parameter we obtain  the equation, 
\begin{equation} \label{eq: NVE_general_1b}
\mathcal{Q}_0 ~ \ddot{\vartheta} + \mathcal{Q}_1 ~ \dot{\vartheta} + \mathcal{Q}_2 ~\vartheta = 0\, ,
\end{equation}
with the $\mathcal{Q}$-factors being equal to:
\begin{equation} \label{eq: NVE_general_1b_factors}
\begin{aligned}
\mathcal{Q}_0 &= 1, &&& \mathcal{Q}_1 &=  E \left.\vphantom{\frac{1}{2}} \left( \frac{\partial_r h_1}{h_1} \right) \right\vert_{r=\bar{r}}, &&& \mathcal{Q}_2 &=\alpha^2_2\, .
\end{aligned}
\end{equation}

We have ended up with two second-order, linear, differential equations. In case that the prefactors turn out not to be rational functions we need to change variables appropriately and make them rational. This small detail aside, both of the NVEs are in a form appropriate for us to examine whether or not they admit Liouville integrable solutions or not. 

In its essence, enforcing differential Galois theory on differential equations amounts to applying the Kovacic's algorithm. The analysis that Kovacic performed is providing the necessary but not sufficient conditions for the classical integrability of differnetial equations of the form described previously, see \cref{eq: NVE_general_1a,eq: NVE_general_1b}. Simply put, in case that these criteria cannot be met, then we can deduce with certainty that there are no Liouville integrable solutions. In turn, this is evidence for the non-integrability of the dynamics in the particular sector under consideration. Since integrability is a property that has to be manifested universally in any given theory, the existence of such a dynamical sector suffices to declare the whole theory as being non-integrable.
	\subsection{The integrable $AdS_3 \times S^3 \times \tilde{S}^3 \times S^1$} \label{sec: ads3s3s3s1}
The point of this section is to demonstrate how the presence of integrability of a given theory is manifested in the method of analytic non-integrability. To do so, we take up the case of string theory in $AdS_3 \times S^3 \times \tilde{S}^3 \times S^1$ which has been shown to be integrable \cite{Babichenko:2009dk}. This particular vacuum is also interesting and relevant for our studies as a special limit of the massive IIA family that we consider in \cref{sec: largesusy}.

The ten dimensional geometry that we consider here is described by the invariant line element 
\begin{equation}
ds^2 = ds^2_{AdS_3} + dr^2 + \frac{1}{\cos^2 \beta} ~ ds^2_{S^3} + \frac{1}{\sin^2 \beta} ~ ds^2_{\tilde{S}^3}\, ,
\end{equation} 
in units where the metric of the AdS space is set to one and we do not write explicitly the flux in the R-R sector as it is not necessary for our purposes. In the above, $\beta$ is just a constant.

The specific submanifolds of the ten dimensional spacetime are given by \cref{eq: general_geometry_2}. We choose a string embedding as in \cref{eq: string_ansatz_1}. 

All of the relations that we derived in the previous sections apply here. We can simply set $h_0=1/h_0=1$, $h_1=1/\cos^2 \beta$ and $h_2=1/\sin^2 \beta$. We derive the following two NVEs 
\begin{equation}
\begin{aligned}
\ddot{\varrho} + (E^2 + \alpha^2_1) \varrho &= 0 \, , \\
\ddot{\vartheta} + \alpha^2_2 \vartheta &= 0\, .
\end{aligned}
\end{equation} 

The above equations admit simple Liouvillian solutions, which are given explicitly in a subsequent section, see \cref{sec: IIA_enhancement}. 

Here we wish to point out that integrability manifests itself in our method through the NVEs having simple Liouville integrable solutions. The same analysis can be performed for the $AdS_3 \times S^3 \times T^4$, which is also known to be integrable \cite{Babichenko:2009dk} as well as the (non)-Abelian T-dual geometries obtained from the above. The $AdS_3 \times S^3 \times T^4$ solution appears as a special limit of the local solution that we examine in \cref{sec: IIB_vacua}. 
\section{Massive IIA with large \texorpdfstring{$\bm{\mathcal{N}=(4,0)$}}{$\mathcal{N}=(4,0)$} superconformal symmetry} \label{sec: largesusy}
While we have provided a step-by-step approach previously in \cref{sec: no_nsns_sector} and in principle we could use all the formulae derived there, we find it useful for illustrational purposes to repeat the procedure here once more explicitly for a specific class of supergravity backgrounds with the warping factors of the metric being explicitly known.

In this section we are dealing with a local solution in massive IIA that preserves $\mathcal{N}=(4,0)$ supersymmetry. It has been shown that from this solution, one is able to construct new global and compact solutions \cite{Macpherson:2018mif}. This was achieved by gluing to the solution a mirror copy of itself in the same vein of \cite{Apruzzi:2013yva}. 

Let us begin by considering the local form of the metric. It is described by the invariant line element \cite{Macpherson:2018mif}
\begin{equation} \label{eq: metric_large_massive_IIA}
ds^2 = \frac{L^2}{\sqrt{h}} ~ ds^2_{AdS_3} + q^2 \sqrt{h} ~ dr^2 + \frac{L^2}{\cos^2 \beta_1 ~ \sqrt{h}} ~ ds^2_{S^3} + \frac{L^2}{\sin^2 \beta_1 ~ \sqrt{h}} ~ ds^2_{\tilde{S}^3}\, ,
\end{equation}
where in the above $L$ and $q$ are constants. The function $h$ is given by 
\begin{equation} \label{eq: rank_function_massive_IIA}
h = F_0 ~ \nu ~ r + c\, ,
\end{equation}
with $c$ being another constant and $F_0$ the Romans' mass.  

The analysis of the supersymmetry conditions yields that there is no NS flux turned on. The metric comes with a non-trivial dilaton and the R-R field, 
\begin{equation} \label{eq: physical_fields_massive_IIA}
\begin{aligned}
e^{-\phi}		&= q ~ h^{5/4}\, , \\
F_4				&= 2 ~ q^2 ~ h \left( L^2 ~ \vol_{AdS_3} + \frac{L^2}{\cos^2\beta_1} ~ \vol_{S^3} + \frac{L^2}{\sin^2\beta_1}~\vol_{\tilde{S}^3} \right)\, ,
\end{aligned}
\end{equation}
and $\beta_1$ is just a constant angle. 

Observe that the metric is of an appropriate form such that it is amenable to the analysis presented for the general metric given by \cref{eq: general_geometry_1}, as the warping factors of the $AdS_3$ part and the $r$-dimension differ only by an uninteresting constant. 

\begin{figure}[H]
	\begin{center}
	\includegraphics[width=9cm]{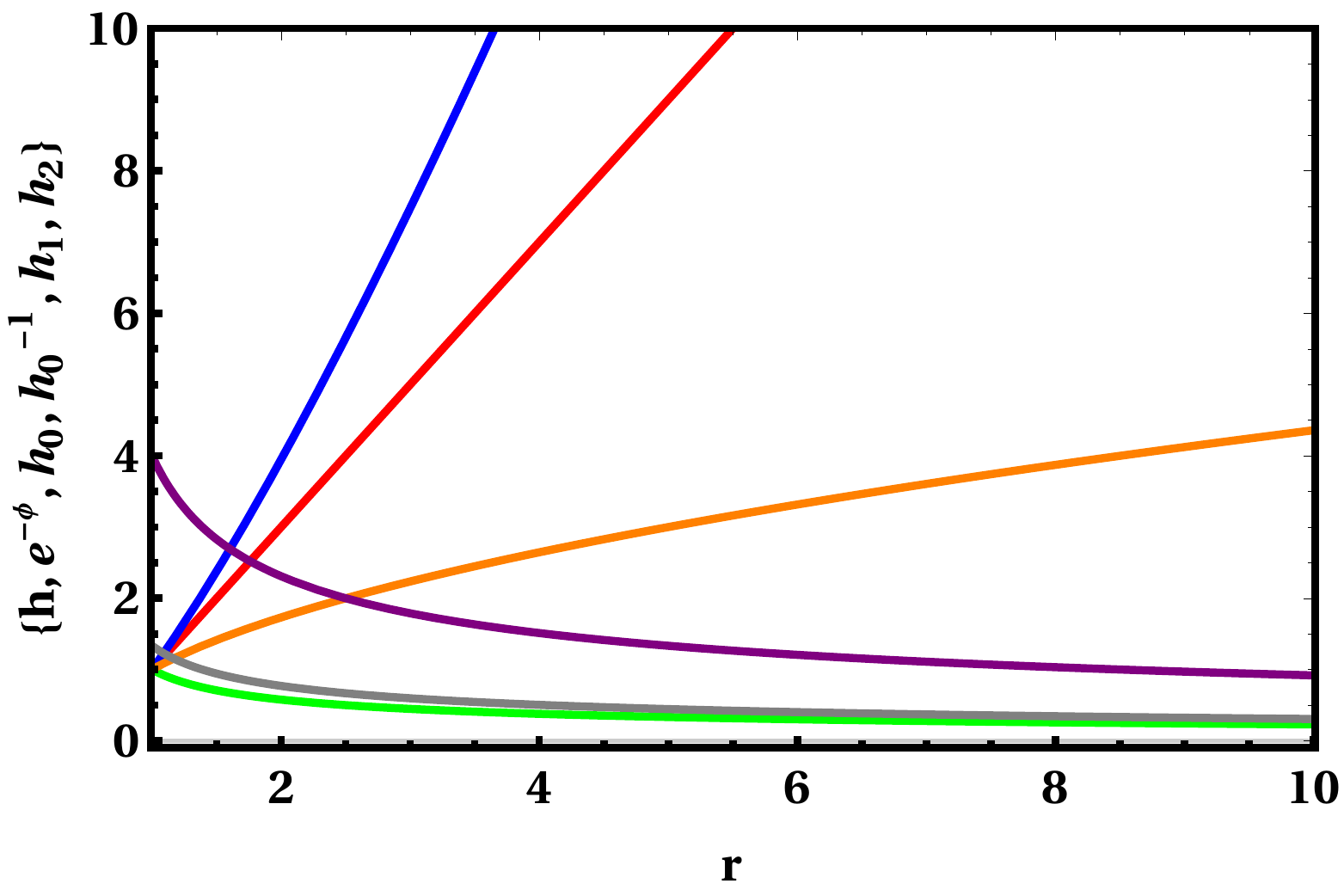} 
	\caption{The $h$-function, the dilaton and the various warp factors of the metric \ref{eq: metric_large_massive_IIA} in the massive IIA theory that realizes large $\mathcal{N}=(4,0)$ superconformal symmetry. We use the language of \cref{eq: general_geometry_1} and we are showing the $h$-function which is the red line, the dilaton $e^{-\phi}$ is te blue line, the green and orange lines are the $h_0$ and $1/h_0$ factors respectively, and finally the $h_1$ and $h_2$ warp factors are represented by the purple and gray lines respectively. The precise numerical values are $F_0=2, \nu = 1, c=-1, L=1$ and $\beta_1 = \pi/3$.}
	\label{large_40_IIA}
  	\end{center}
\end{figure}

In order to gain a better physical understanding and a solid grasp of this particular class of models, we plot the linear function $h$, the various warp factors of the metric, as well as the dilaton in \cref{large_40_IIA} 

If we consider the limit of the metric in which the Romans' mass vanishes, $F_0=0$, then the solution returns to being the standard $AdS_3 \times S^3 \times \tilde{S}^3 \times S^1$, which preserves $\mathcal{N}=(4,4)$ supersymmetry. Observe that in this limit none of the physical fields in the above supergravity description \cref{eq: metric_large_massive_IIA,eq: rank_function_massive_IIA,eq: physical_fields_massive_IIA} depend on the parameter $\nu$ which ties really well with the fact that there is an enhancement of supersymmetry.  

The local massive IIA solution presented above has an interpretation as either D$8$-branes, or O$8$-planes, or both that have backreacted on the geometry. In the $F_0 \neq 0$ case the warping factor depends on the constant $\nu$ and hence the backreacting D$8$/O$8$ setup supports only $\mathcal{N}=(4,0)$.

In the non-trivial Romans' mass case, $F_0 \neq 0$, the internal space of the above supergravity configuration \cref{eq: metric_large_massive_IIA} is not compact. By assuming that $F_0 >0$ and setting $\nu=1$, the interval is bounded from below at a distance $r=-c/F_0$. This picture is consistent with the D$8$/O$8$ setup identification that wraps the $AdS_3 \times S^3 \times \tilde{S}^3$. Be that as it may, the interval is still not bounded from above (as $r \rightarrow \infty$). This is an issue, however, that has been addressed and resolved in \cite{Macpherson:2018mif}. 

A compact solution was obtained by gluing together a second copy (a mirror copy) of \cref{eq: metric_large_massive_IIA} to itself, utilizing the method deveoped in \cite{Apruzzi:2013yva}. At the point where the two copies of the local solution above connect, there should be a D$8$-brane defect such that the Romans' mass jumps, while the metric and the dilaton are kept being continuous. The simplest arrangement that achieves that is to consider that the gluing occurs at $r=0$ and the Romans' mass changes sign from positive to negative as we pass from the negative region of the $r$-coordinate to the positive one. More concretely, we take the warp factor to be piecewise defined via 
\begin{equation}
h = \begin{dcases}
        c + |F_0|~r & r < 0\, , \\
   		c - |F_0|~r & r > 0\, , \\
    \end{dcases}
\end{equation}
and the $r$-dimension becomes an interval $\mathcal{I}_r$ that is bounded between two D$8$/O$8$ configurations at the points $r = \pm c/|F_0|$. 

Before delving into the main computational steps that are necessary in order to determine the Liouville (non)-integrability, some final brief comments are in order. To begin with, there is absolutely no necessity to glue the two mirrored copies of the local solutions together at $r=0$. This just provides the  easiest alternative.  In principle, one can consider of gluing together an arbitrary number of the mirrored copy at each point of the interesection along the lines of \cite{Cremonesi:2015bld}. More precisely by doing so, one could construct an infinite number of completely distinct global solutions. Also, to the extend of our knowledge there is no explicit intersecting brane realization of the local solution presented here, though it would be very interesting. Furthermore, after creating the global solution the quantization of the charges can be performed without any issues. Lastly, the dual superconformal field theory realization of this supergravity vacuum is conjecturally the addition of matter appropriately in the $AdS_3 \times S^3 \times \tilde{S}^3 \times S^1$ field theory dual, such that the $\mathcal{N}=(4,4)$ supersymmetry is broken down to $\mathcal{N}=(4,0)$. The above matters are more thoroughly discussed in \cite{Macpherson:2018mif}, where the charge quantization is also performed.

We will be considering the simple form for the $h$-function given by \cref{eq: rank_function_massive_IIA} and consequently all the warp factors of the metric for simplicity. Their piecewise character is known to hold. In other words, as we are dealing with the local form of the metric the conclusions that we reach can be assumed to hold true in any interval of the compact global supegravity solution.  

We begin by evaluating explicitly the lagrangian associated to the $\sigma$-model action \cref{eq: polyakov_action} for a string soliton of the form \cref{eq: string_ansatz_1} for the class of backgrounds defined by the local metric \cref{eq: metric_large_massive_IIA}. The computation yields 
\begin{equation}
\mathcal{L} = \frac{L^2}{\sqrt{h}} ~ (\cosh^2 \rho ~ \dot{t}^2 - \dot{\rho}^2 + \sinh^2 \rho ~ \alpha^2_1) - q^2 ~ \sqrt{h} ~ \dot{r}^2 - \frac{L^2}{\cos^2 \beta_1 ~ \sqrt{h}} (\dot{\theta_1}^2 - \sin^2 \theta_1 \alpha^2_2)\, .
\end{equation}
From the above we can obtain the equations of motion for the various coordinates. The one related to the time direction is simple and we integrate it immediately. All of the equations read 
\begin{equation} \label{eq: eom_massiveIIA_large_susy}
\begin{aligned}
\dot{t}			&= \frac{E~\sqrt{F_0~\nu~r+c}}{\cosh^2 \rho} \, , \\
\ddot{\rho}		&= \frac{F_0~\nu}{2(F_0~\nu~r+c)} ~ \dot{r} ~ \dot{\rho} + \frac{1}{2}\sinh 2\rho \left(\frac{E^2 (F_0 ~ \nu ~ r + c)}{\cosh^4\rho} + \alpha^2_1  \right) \, ,\\
\ddot{r}		&= - \frac{F_0~\nu}{4(F_0~\nu~r+c)} ~ \dot{r}^2 + \frac{1}{4} \left( \frac{L^2~F_0~\nu}{q^2~(F_0~\nu~r+c)^2}~\left( \frac{E^2(F_0~\nu~r+c)}{\cosh^2 \rho} + \sinh^2 \rho \alpha^2_1 - \dot{\rho}^2 \right) \right. \\
&\left.-\frac{L^2~F_0~\nu}{q^2(F_0~\nu~r+c)^2} ~ \frac{1}{\cos^2 \beta_1} ~ (\dot{\theta_1}^2 - \sin^2 \theta_1 \alpha^2_2)
 \right)\, , \\
\ddot{\theta_1}	&= \frac{F_0 ~ \nu}{2(F_0~\nu~r+c)}~\dot{r}~\dot{\theta_1} - \cos \theta_1 ~ \sin \theta_1 ~ \alpha^2_2\, ,
\end{aligned}
\end{equation}
where we have subsituted the equation of motion for the time-component, $t(\tau)$, in the rest.

The equations of motion given above are constrained by the Virasoro conditions. For the class of backgrounds of this section, the constraints are 
\begin{equation} 
\begin{aligned}
2 T_{\tau \tau} = 2 T_{\sigma \sigma} &=  \frac{L^2}{\sqrt{h}} ~ (-\cosh^2 \rho ~ \dot{t}^2 + \dot{\rho}^2 + \sinh^2 \rho ~ \alpha^2_1) + q^2 \sqrt{h} ~ \dot{r}^2 \\ 
&+ \frac{L^2}{\cos^2 \beta_1 ~ \sqrt{h}} ~ (\dot{\theta_1}^2 + \sin^2 \theta_1 \alpha^2_2)\, , \\
T_{\tau \sigma} = T_{\sigma \tau} &= 0 \, .
\end{aligned}
\end{equation}

Of course, in this case as well, the energy-momentum tensor is conserved -$\nabla^{\alpha}T_{\alpha \beta}=0$- when we enforce the equations of motion derived from the $\sigma$-model lagrangian, \cref{eq: eom_massiveIIA_large_susy}.  

The invariant plane of solutions is obviously given by 
\begin{equation} \label{eq: gen_invariant_massiveIIA}
\rho = \dot{\rho} = \ddot{\rho} = \theta_1 = \dot{\theta_1} = \ddot{\theta_1} = 0\, ,
\end{equation}
alongside with the simple solution
\begin{equation} \label{eq: general_simple_sltn_r_masisveIIA}
r = \bar{r} = E ~ \tau\, .
\end{equation}

To avoid dealing with the complicated coupled equations of motion, we consider fluctuations of the string around this plane of solutions and on the simple solution for the $r$-cooridnate given by \cref{eq: general_simple_sltn_r_masisveIIA}. 

Let us first consider the NVE associated to the $\theta_1$-direction. We fluctuate as $\theta_1 = 0 + \varepsilon~\vartheta$ in the limit $\varepsilon \rightarrow 0$ and work only up to linear order in the small parameter. This yields the equation 
\begin{equation} \label{eq: NVE_theta1_massiveIIA_large}
\ddot{\vartheta} - \frac{L~F_0~\nu~E}{2(L~F_0~\nu~E~\tau+c~q)}~\dot{\vartheta} - \alpha^2_2 ~ \vartheta = 0\, .
\end{equation}

The above equation admits analytic solutions in terms of Bessel functions of the first ($J_n(z)$) and the second-kind ($Y_n(z)$). Explicitly the solution can be written as 
\begin{equation}
\vartheta = \mathcal{P}(\tau)~\left[\vphantom{\frac{1}{2}} c_1 ~ J_{3/4}\left(\alpha_2 \left(\tau + \frac{c~q}{L~F_0~\nu~E} \right) \right) + c_2 ~ Y_{3/4}\left(\alpha_2 \left(\tau + \frac{c~q}{L~F_0~\nu~E} \right) \right) \right] \, ,
\end{equation}
with $c_{1,2}$ the two constants of integration and 
\begin{equation}
\mathcal{P}(\tau)= (c~q+L~F_0~\nu~E~\tau)^{3/4}\, .
\end{equation}

Bessel functions are non-Liouvillian for general values of their ranks. However, these functions are clasically integrable for half-integer rank as they can be equivalently expressed in terms of polynomials and trigonometric functions. This is the case in the above solution. 

Having solved one of the NVEs with a Liouvillian function, the chance of spotting a non-integrable behaviour lies in the NVE for the $\rho$-coordinate which we derive now. To that end, we expand as $\rho = 0 + \varepsilon~\varrho$ in the small $\varepsilon$-limit as above and evaluated on the invariant plane of solutions \cref{eq: gen_invariant_massiveIIA} and the simple solution \cref{eq: general_simple_sltn_r_masisveIIA}. A striaghtforward calculation yields
\begin{equation} \label{eq: NVE_rho_massiveIIA_large}
\ddot{\varrho} - \frac{L~F_0~\nu~E}{2(c~q+L~F_0~\nu~E~\tau)} ~ \dot{\varrho} + \left(E^2 ~ \left(c + \frac{L~F_0~\nu~E~\tau}{q} \right) + \alpha^2_1 \right)~\varrho = 0 \, .
\end{equation}

The above equation does not admit simple analytic Liouvillian solutions. This, however, does not mean that within this class of supegravity vacua none of it is integrable. It means that the full family is non-integrable for general values of the parameters. 

Therefore, we have ultimately spotted a particular sector of the string soliton that exhibits non-integrable dynamics. This is enough to declare that the supergravity vacuum descirbed by \cref{eq: metric_large_massive_IIA} and of course the associated global supergravity solution obtained after gluing mirror copies together are non-integrable. Though the boundary superconformal field theory description is still elusive, by virtue of the AdS/CFT and the discussion in \cref{sec: field_theory_comments} the non-integrability property of this class of supergravity backgrounds should be shared with the holographic field theory description as well. 
We can perform a more extensive check, however, by examining an equivalent form of this equation as explained in \cref{app: analytic_kovacic}. In \cref{sec: largesusy_app} we prove that no possible tuning of the parameters can lead to a clasically integrable solution.

On top of the above, we performed a more exhaustive check by examining an equivalent form of this equation as explained in \cref{app: analytic_kovacic}. In \cref{sec: largesusy_app} we prove that no possible tuning of the parameters can lead to a clasically integrable solution. To do so, we enforce eac hand every step of the Kovacic algorithm on the differential equation and we derive the said conclusion -namely the non-integrability- as a final result of the algorithmic process. Hence \textit{each supergravity vacuum that can be constructed from the above class of solutions and respects the string theory constraints that we discussed is non-integrable in the Liouvillian sense}. 

Note, however, that within the non-integrable theories there might be one or more integrable sub-sectors. 
	\subsection{Comments on the low-energy description of the modes}
In this section we want to discuss very briefly the limit in which the wrapping of the string soliton along the cyclic cooridnates is zero. We start by taking the two NVEs derived above and setting directly $\alpha_1=\alpha_2=0$. Recall that the NVE for the $\theta_1$-coordinate has already a Liouville integrable solution even at the level of the extended string, while for the NVE of the $r$-dimension no such solution exists. 

Applying the zero wrapping of the string limit in \cref{eq: NVE_theta1_massiveIIA_large} yields 
\begin{equation}
\ddot{\vartheta} - \frac{L~F_0~\nu~E}{2(c~q+L~F_0~\nu~E~\tau)}~\dot{\vartheta} = 0\, ,
\end{equation}
which clearly admits integrable solutions. While the solution to the above equation is not essential for our studies we give it explicitly below for completeness 
\begin{equation}
\vartheta = c_1 + c_2~ \frac{(2~c~q+2~L~F_0~\nu~E~\tau)^{3/2}}{3~L~F_0~\nu~E}\, ,
\end{equation}
with $c_{1,2}$ being the constants of integration. Now, we turn our attention to the NVE pertaining to the $\rho$-dimension which is given by \cref{eq: NVE_rho_massiveIIA_large} and again we set $\alpha_1=\alpha_2=0$ to obtain
\begin{equation}
\ddot{\varrho} - \frac{L~F_0~\nu~E}{2(c~q+L~F_0~\nu~E~\tau)} ~ \dot{\varrho} + E^2 ~ \left(c + \frac{L~F_0~\nu~E~\tau}{q} \right) ~\varrho = 0 \, ,
\end{equation}
which can be solved by 
\begin{equation}
\varrho = c_1 \cos \mathcal{A}(\tau) + c_2 \sin \mathcal{A}(\tau)\, ,
\end{equation}
with 
\begin{equation}
\mathcal{A}(\tau) = \frac{2 (c~q+L~F_0~\nu~E~\tau)^{3/2}}{3~L~F_0~\sqrt{q}~\nu}\, .
\end{equation}
The above solution is Liouvillian. At this point though we should be careful and not rush to declare the supegravity description as being integrable for the point-like limit of the string. This might be the case as well, however the fact that both of the NVEs above have classically integrable is not conclusive evidence for the integrability of the vacuum under consideration. It is perhaps suggestive and motivating evidence to look for a formal proof of integrability in this limit. 

Another comment is also in order regarding the results in this section. It appears, once more, that the classical integrability for the extended string motion in a given background is a much more stringent and restrictive requirement than the integrability of point particles. 
	\subsection{Comments on the $\mathcal{N}=(4,4)$ limit; vanishing Romans' mass} \label{sec: IIA_enhancement}
As we have discussed the method we have adopted in this work cannot prove the integrability of a system. Be that as it may, if a system is integrable then this should be manifested in the NVEs are well and they should yield Liouvillian solutions. 

As we have already mentioned above, there is a special limit of the local supergravity solution that we considered here. This is the limit $F_0=0$. Upon taking this limit, the metric returns to the familiar $AdS_3 \times S^3 \times \tilde{S}^3 \times S^1$ solution which preserves $\mathcal{N}=(4,4)$ supersymmetry and, as we already mentioned previously, is known to be classically integrable \cite{Babichenko:2009dk}. This fact should be reflected in the NVEs that we derived in this section upon taking the $F_0=0$ limit. We shall show that this indeed is the case.

As a consistency check, here we consider the said limit in the NVEs we derived above, see \cref{eq: NVE_theta1_massiveIIA_large,eq: NVE_rho_massiveIIA_large}. We begin with the NVE for the $\theta_1$ dimension. In the limit we consider here it becomes, 
\begin{equation}
\ddot{\vartheta} + \alpha^2_2 \vartheta = 0\, , 
\end{equation}
which has a very simple solution 
\begin{equation}
\vartheta = c_1 ~ \cos (\alpha_2 ~ \tau) + c_2 ~ \sin (\alpha_2 ~ \tau) \, .
\end{equation}
The above is Liouville integrable. 

Now, we turn our attention to the NVE of the $r$-dimension and we evaluate it in the $F_0=0$ limit. It reads 
\begin{equation}
\ddot{\varrho} + (c~E^2+\alpha^2_1) ~ \varrho = 0 \, .
\end{equation}
The above admits a Liouvillian solution as well, given by 
\begin{equation}
\varrho = c_1~e^{\tau~\sqrt{\mathcal{B}}} + c_2~e^{-\tau~\sqrt{\mathcal{B}}} \, ,
\end{equation}
with $\mathcal{B}$ being given by 
\begin{equation}
\mathcal{B} = -c~E^2-\alpha^2_1 \, .
\end{equation}

We, again, feel necessary to stress that the above does not constitue a proof of integrability for the $AdS_3 \times S^3 \times \tilde{S}^3 \times S^1$ supergravity description. It rather serves as a neat consistency check of our equations. 
	\subsection{Comments on instantonic string configurations} \label{sec: insta_IIA}
One might ponder upon the possibility of finding integrable solutions for the instantonic string mode $E=0$. We go back to the NVEs that we derived \cref{eq: NVE_theta1_massiveIIA_large,eq: NVE_rho_massiveIIA_large} and we set directly $E=0$ to obtain
\begin{equation}
\begin{aligned}
\ddot{\varrho}		&= 	-\alpha^2_1 ~ \varrho	\, ,\\
\ddot{\vartheta}	&=	-\alpha^2_2 ~ \vartheta	\, ,
\end{aligned}
\end{equation}
both of which have obvious Liouvillian solutions. The system of the NVEs has reduced to the description of two decoupled simple harmonic oscillators. 

Be that as it may, integrability is a universal property of a given theory or a given sector for that matter. Hence, since the $E \neq 0$ strings have non-integrable dynamics, the string sector is unequivocally non-integrable. The above result might be a suggestion that the instantonic configurations as a subsector of the theory possesses integrable dynamics, but this requires a separate and very careful treatment and goes beyond the scope of this work. We felt that this discussion was necessary, as the solution $E=0$ will appear after going through the Kovacic algorithm explicitly, see \cref{sec: largesusy_app}.    

All in all, this solution does not change anything with our characterization as the class of supergravity backgrounds in \cref{sec: largesusy} as non-integrable.
\section{IIB with small \texorpdfstring{$\bm{\mathcal{N}=(4,0)$}}{$\mathcal{N}=(4,0)$} superconformal symmetry} \label{sec: IIB_vacua}
This section is devoted to another class of supergravity solutions that was also obtained in \cite{Macpherson:2018mif}. The local supergravity solution that will concern us here is a solution in type IIB that preserves small $\mathcal{N}=(4,0)$. Since we have worked out explicitly all the necessary steps and relations in the previous sections, here we will move a bit more quickly towards our final goal. 

The local solution that concerns us in this section is given by
\begin{equation} \label{eq: local_IIB_small}
ds^2 = L^2 \left(\frac{1}{\sqrt{h}}~ds^2_{AdS_3} + \sqrt{h}~dr^2 +  r^2~\sqrt{h}~ds^2_{S^3} + \frac{1}{\sqrt{h}}~ds^2_{\tilde{S}^3} \right) \, ,
\end{equation}
which comes with a dilaton field and a non-trivial three-form in the R-R sector. They are equal to 
\begin{equation}
\begin{aligned}
e^{-\phi}	&= \frac{c_2}{2~L}~\sqrt{h}\, , \\
F_3			&= c_2~\left( \vol_{AdS_3} + \vol_{\tilde{S}^3} \right) + c_1~\nu~\vol_{S^3} \, ,
\end{aligned}
\end{equation} \label{eq: def_h_IIB_small}
and the function $h$ is given by:
\begin{equation}
h	= a + \frac{c_1}{c_2~r^2} \, .
\end{equation}

As in the previous class of backgrounds, here we also do not have an NS-NS sector as it vanishes due to supersymmetry consistency considerations. Note that the metric \cref{eq: local_IIB_small} is of a form appropriate to use directly the analysis presented for the general metric given by \cref{eq: general_geometry_1}. Before turning our attention to the NVEs and spotting  a non-integrable sector some comments are necessary in order to get a better understanding of the local solution here. 

If we consider the constant $a$ that appears in the definition of the function $h$ \cref{eq: def_h_IIB_small} to take the value $a=1$ and then consider appropriate values for the other two constants $c_{1,2}$ such that $c_1 \cdot c_2 > 0$, then $h$ becomes the warp factor of a D$5$-brane. Therefore, the natural interpretation is that the solution presented here portrays D$5$ branes that backreact on $AdS_3 \times S^3 \times \mathbb{R}^4$. The latter is not a compact solution. 

It is also quite clear that upon taking the limit $c_1=0$, none of the physical fields of the local solutions depends on the parameter $\nu$, and we have a supersymmetry enhancement to $\mathcal{N}=(4,4)$. This picture is consistent with the fact that in this limit the local solution becomes $AdS_3 \times S^3 \times T^4$. 

As it has been pointed out in \cite{Macpherson:2018mif}, one might be able perform the gluing procedure as was done for the massive IIA class of backgrounds that we considered in \cref{sec: largesusy} and obtain a global compact solution. However, this is not necessary in this type IIB example as there exists an easier way to achieve a well behaved global solution. 

We can tune the parameters in such a way that $a<1$. This immediately bounds the $r$-dimension to an interval $\mathcal{I}_r$ taking values in the region 
\begin{equation}
\left[0, \sqrt{\frac{c_1}{|a|~c_2}} \right]\, ,
\end{equation}
and the solution, now, becomes compact. 

The above can become even more pellucid if one considers the coordinate transformation 
\begin{equation}
r \rightarrow \sqrt{\frac{c_1}{|a|~c_2}}~\cos r \, ,
\end{equation} 
alongside with the coordinate redefinition 
\begin{equation}
L^2 \rightarrow |a|~L^2 \, .
\end{equation}
The above shifts modify the local metric solution and the dilaton, while leaving the three-form in the R-R sector completely the same. As the metric now depends explicitly on the constant $c_1$ it can no longer be set to zero and hence only $\mathcal{N}=(4,0)$ superconfomral symmetry is generically preseved. The newly defined internal radius $r$ is now bounded in the region $r \in (0,\pi/2)$. The two endpoints of the said region correspond to singularities, which, however, have a physical explanation. Close to the $r \rightarrow 0$ lower limit, the metric becomes that of an O$5$ orientifold plane that is wrapped around the $AdS_3 \times \tilde{S}^3$ part of the metric. Taking a closer look at the upper limit, $r \rightarrow \pi/2$, one sees that the local solution becomes that of a D$5$-brane that wraps the $AdS_3$ part of the metric and either $S^3$ or $\tilde{S}^3$. 

More details related on the above discussion can be found in \citep{Macpherson:2018mif} and we urge the interested reader to follow the said discussion. For our purposes, we wanted to understand that the above solution defines a compact solution that contains D$5$-branes and O$5$ orientifold planes that wrap appropriate subregions of the metric. We will be working with the metric in the form \cref{eq: local_IIB_small} though it requires some fine tuning of the constants, as for our purposes this not a limitation\footnote{of course one is able to reach the sae conclusions using the metric obtained after the coordinate change and the corresponding redefinition. The form of the metric we used is more convenient here as the general formuale of \cref{sec: no_nsns_sector} are directly applicable.}. As a final comment, we want to mention that in this example, as well as in the previous one in the massive IIA, the flux quanitzation has been explicitly performed. The pucnhline is that the local solution that we consider in this section defines a prime example of a supergravity vacuum to a two-dimensional superconformal field theory with small $\mathcal{N}=(4,0)$ that has not yet been determined to the best of our knowledge. 

\begin{figure}[H]
	\begin{center}
	\includegraphics[width=9cm]{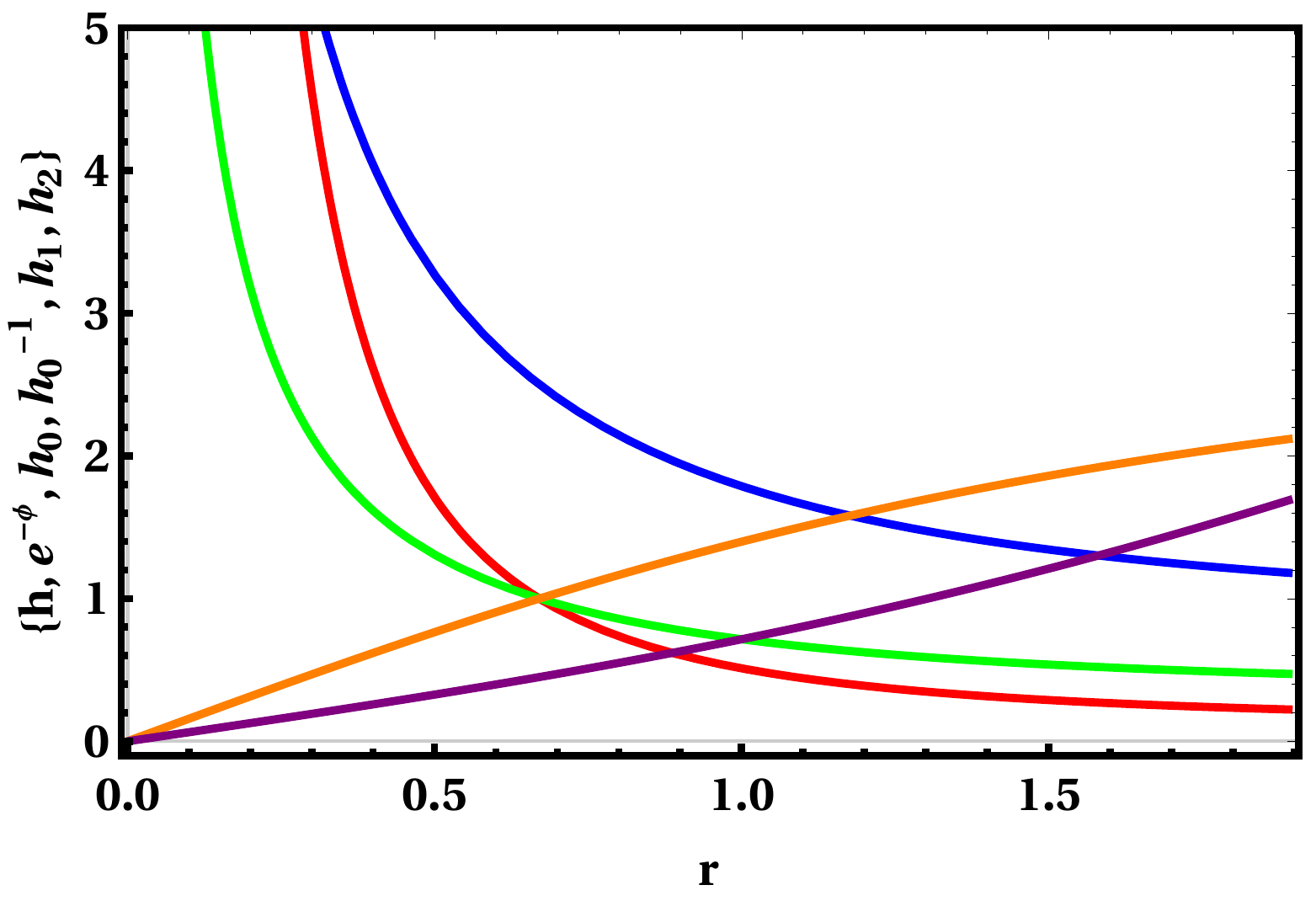} 
	\caption{The function, $h$, which is the red line and the various warp factors of the metric \cref{eq: local_IIB_small} in type IIB theory that realizes small $\mathcal{N}=(4,0)$ superconformal symmetry. The dilaton, $e^{-\phi}$ is the blue line, the warping factors $h_0$ and $h_2$ are equal and denoted by the orange line, the $h_1$-factor is the purple and finally the $1/h_0$ is depicted in green. The parameters are chosen to be $a=1/9$, $L=1$, $c_1=2$ and $c_2=5$.}
	\label{small40IIB}
  	\end{center}
\end{figure}

In order to have a better feel of the solution with which we are working here, we plot the various factors that enter the description in \cref{small40IIB}. 

We are now ready to examine the (non)-integrability of the above class of supergravity solutions. Since we have already performed the analysis in terms of general function in \cref{sec: no_nsns_sector} and also in \cref{sec: largesusy} for an example with specific warp factors, here we are content with applying the special values of the local solution to the general formulae that we derived in \cref{sec: no_nsns_sector}. It is a very straightforward exercise to do so, and we begin by examining the NVE for the $\theta_1$ coordinate. We use the equations given by \cref{eq: NVE_general_1b,eq: NVE_general_1b_factors} evaluated for our particular supergravity vaccum. We obtain, after a straightforward evaluation,
\begin{equation} \label{eq: NVE1_IIB}
\ddot{\vartheta} + \frac{1}{\tau}~\frac{c_1 + 2~a~c^2_2~E^2~\tau^2}{c_1 + a~c^2_2~E^2~\tau^2}\dot{\vartheta} + \alpha^2_2 \vartheta = 0 \, .
\end{equation}
The above solution does not admit simple analytic Liouvillian solutions. 

At this point we can declare the class of supergravity solutions described by \cref{eq: local_IIB_small} to be non-integrable in general, since we have spotted a particular sector of the string soliton that exhibits non-integrable dynamics. Though the boundary superconformal field theory description is still elusive, by virtue of the AdS/CFT and the discussion in \cref{sec: field_theory_comments} the non-integrability property of this class of supergravity backgrounds should be shared with the holographic field theory description as well. 

For completeness we will also present the NVE asosciated with the $\rho$-dimension, though it is not of substantial use at this point. Again, a straightforward computation of the general relation given by \cref{eq: NVE_general_1a,eq: NVE_general_1a_factors} we arrive at 
\begin{equation} \label{eq: NVE2_IIB}
\ddot{\varrho}  + \frac{1}{\tau}~\frac{c_1}{c_1+a~c^2_2~E^2~\tau^2}~\dot{\varrho} + \left(\frac{c_1+c_2~\alpha^2_1~\tau^2+a~c^2_2~E^2~\tau^2}{c_2~\tau^2} \right)~\varrho = 0 \, .
\end{equation}
The discussion for the solution of the above differential equation is the same as the in the case of the $\theta_1$ NVE above.

However, as we have already discussed, the failure of Kovacic’s algorithm here is a mere statement that not all every supergravity  background that is allowed in this class of solutions is integrable. It is not a conclusive statement that there are no integrable vacua, amongst the whole class. Since, the NVEs are parameterized in terms of some constants, we should formally apply the full power of the analytic Kovacic algorithm by hand and examine each and every step. 

This is what we did. Since following the steps of the Kovacic algorithm does not provide any physical intution, though it is a very elegant mathematical statement, we chose to present the necessary steps of the said computations in \cref{sec: smallsusy_app}. Here we just discuss the implications. 

We examined all the steps of the two cases of the Kovacic algorithm that the NVE of the $\rho$-dimension satisfies. If there were a Liouvillian solution, the procedure would return the desired answer and it would fix the particular values of the parameters for which this occurs. However, we showed that there is no such combination (up to the special limits that we discuss explicitly below), and hence \textit{every supergravity vacuum that can be constructed from the above class of solutions and respects the string theory constraints that we discussed is non-integrable in the Liouvillian sense}.
	\subsection{Comments on the low-energy description of the modes} \label{sec: point_like_IIB}
As we did for the massive IIA class of solutions, here we deal very briefly with the limit in which the wrapping of the string soliton along the cyclic cooridnates is zero. We start by taking the two NVEs derived above, \cref{eq: NVE1_IIB,eq: NVE2_IIB}, and setting directly $\alpha_1=\alpha_2=0$.

The NVE for the $\theta_1$ dimension becomes 
\begin{equation}
\ddot{\vartheta} + \frac{1}{\tau}~\frac{c_1 + 2~a~c^2_2~E^2~\tau^2}{c_1 + a~c^2_2~E^2~\tau^2}\dot{\vartheta} = 0 \, ,
\end{equation}
which admits Liouvillian solutions in terms of a hyperbolic trigonometric function. Specifically its solution is  given by
\begin{equation}
\vartheta = d_1 - \frac{d_2}{\sqrt{c_1}}~\arctanh\left(\sqrt{\mathcal{S(\tau)}} \right) \, ,
\end{equation}
with $d_{1,2}$ the constants of integration and the function $\mathcal{S}(\tau)$ is equal to
\begin{equation} \label{eq: mathcal_S_IIB}
S(\tau) = 1 + \frac{a~c^2_1~c^2_2~\tau^2}{c_1} \, .
\end{equation}

Now, we shift our focus to the NVE for the $\rho$-coordinate. We take the point-like string limit on the equation \cref{eq: NVE2_IIB} and obtain 
\begin{equation} 
\ddot{\varrho}  + \frac{1}{\tau}~\frac{c_1}{c_1+a~c^2_2~E^2~\tau^2}~\dot{\varrho} + \left(\frac{c_1 + a~c^2_2~E^2~\tau^2}{c_2~\tau^2} \right)~\varrho = 0 \, .
\end{equation}
The equation given above admits the solution
\begin{equation}
\begin{aligned}
\varrho &= d_3~\cos\left(\frac{\sqrt{\mathcal{R}(\tau)} - \sqrt{c_1}~\arctanh\sqrt{\mathcal{S}(\tau)}}{\sqrt{c_2}} +  \right) \\
&+ d_4~\sin\left(\frac{\sqrt{\mathcal{R}(\tau)} - \sqrt{c_1}~\arctanh\sqrt{\mathcal{S}(\tau)}}{\sqrt{c_2}} \right) \, ,
\end{aligned}
\end{equation}
where in the above $d_{3,4}$ are just constants, $\mathcal{S}(\tau)$ is given by \cref{eq: mathcal_S_IIB} and the function $\mathcal{R}(\tau)$ is 
\begin{equation}
\mathcal{R}(\tau) = c_1 + a~c^2_2~E^2~\tau^2\, .
\end{equation}
The solution presented above is Liouvillian as well. 
	\subsection{Comments on the $\mathcal{N}=(4,4)$ enhancement} \label{sec: IIB_enhancement}
There is another very special limit that we can consider which is of interest. This is the limit $c_1=0$ where the metric returns to beign locally $AdS_3 \times S^3 \times T^4$ and supersymmetry is enhanced to $\mathcal{N}=(4,4)$. As we already mentioned, it is known that the $AdS_3 \times S^3 \times T^4$ supergravity descirption is classically integrable \cite{Babichenko:2009dk}. This fact should be reflected in the NVEs that we derived in this section upon taking the $c_1=0$ limit. We shall show that this indeed is the case. This serves as a nice consistency check of the equations that we derived. 

We consider the said limit in the NVEs we derived above, see \cref{eq: NVE1_IIB,eq: NVE2_IIB}. We begin with the NVE for the $\theta_1$ dimension. In the $c_1=0$ limit that we consider here the NVE becomes,
\begin{equation}
\ddot{\vartheta} + \frac{2}{\tau}~\dot{\vartheta}+\alpha^2_2~\vartheta = 0\, ,
\end{equation} 
which admits the Liouvillian solution 
\begin{equation}
\vartheta = \frac{e^{-i~\alpha_2~\tau}}{2~\tau}~\left(2~g_1 - i~g_2~\frac{e^{2~i~\alpha_2~\tau}}{\alpha_2} \right)\, ,
\end{equation}
with $d_{1,2}$ being constants.

We proceed to the examination of the NVE derived for the $\rho$-coordinate. Upon taking the $c_1=0$ limit it becomes,
\begin{equation}
\ddot{\varrho} + (\alpha^2_1 + a~c_2~E^2) \varrho = 0\, ,
\end{equation}
which admits a Liouville integrable solution as well, given by 
\begin{equation}
\varrho = f_1~e^{\tau~\sqrt{\mathcal{C}}} + f_2~e^{-\tau~\sqrt{\mathcal{C}}}
\end{equation}
with $f_{1,2,}$ the constants of integration, while $\mathcal{C}$ is given by 
\begin{equation}
\mathcal{C} = -\alpha^2_1 - a~c_2~E^2\, .
\end{equation}

We, once more, feel that it is necessary to point out that the above does not constitue a proof of integrability for the $AdS_3 \times S^3 \times T^4$ supergravity description. It rather serves as a nice illustrative example suggesting that the integrability of a given background has to be manifested in the NVEs associated with the said background and is a consistency check of our equations as well. 
	\subsection{Comments on instantonic string configurations}
As we did in the previous class of vacua, here we examine the zero-energy strings as well. We use the NVEs that we derived for the $\theta_1$ and the $\rho$ dimensions, \cref{eq: NVE1_IIB,eq: NVE2_IIB}, and set $E=0$. The newly obtained ones read 
\begin{equation}
\begin{aligned}
&\ddot{\vartheta} + \frac{1}{\tau} ~ \dot{\vartheta} + \alpha^2_2~\vartheta = 0\, ,\\
&\ddot{\varrho} + \frac{1}{\tau} ~ \dot{\varrho} + \left(\alpha^2_1 + \frac{c_1}{c_2}\frac{1}{\tau^2} \right)~\varrho = 0 \, .
\end{aligned}
\end{equation}
The above equations admit analytic solutions in terms of Bessel functions of the first and the second, which are given below for completeness 
\begin{equation}
\begin{aligned}
\vartheta 	&= m_1~J_{0}(\alpha_2~\tau) + m_2~Y_{0}(\alpha_2~\tau)\, ,\\
\varrho 	&= m_3~J_{i~\sqrt{\frac{c_1}{c_2}}}(\alpha_1~\tau) + m_2~Y_{i~\sqrt{\frac{c_1}{c_2}}}(\alpha_1~\tau) \, ,
\end{aligned}
\end{equation}
where in the above $m_{1,2,3,4}$ are constants. 

As we have already discussed, Bessel functions are integrable only for half-integer rank, and hence the instantonic strings here do not admit Liouvillian solutions as opposed to the massive IIA picture in \cref{sec: insta_IIA}.
\section{Exceptional \texorpdfstring{$\bm{\mathcal{N}=8$}}{$\mathcal{N}=8$} superconformal symmetry in massive IIA} \label{sec: exceptional}
In this section we examine the (non)-integrability of a local supergravity solution that realizes the exceptional $F(4)$ superalgebra. There exists a unique solution for that particular choice of the algebra, a result that has been obtained in \cite{Dibitetto:2018ftj}. In that work, the authors managed to derive the particular supergravity solution that will concern as here in a twofold way; by utilizing the near-horizon limit of a known brane solution and by considering the analysis and implications of the supersymmetry conditions. 

The brane solution that manages to give an appropriate metric that realizes the eceptional superalgerba of interest is a stack of D$2$-branes that extend in the $\{0,1,2\}$ subspace of the ten-dimensional geometry and an O$8$ orientifold plane that extends along all the coordinates except for the $x^2$-dimension. The near-horizon limit of the brane intersection that we mentioned can be depicted by the following schematic representation, see \cref{table: D2_O8} 
\begin{table}[H]
\begin{center}
\begin{tabular}{ |c|c|c|c|c|c|c|c|c|c|c|c|}
 \hline
 &&&&&&&&&&\\[-0.95em] 
   									& $x^0$ & $x^1$ & $x^2$ & $x^3$ & $x^4$ & $x^5$ & $x^6$ & $x^7$ & $x^8$ & $x^9$			\\ 
 \hline
 D$2$-brane 	& --- & --- & --- & $\bullet$ & $\bullet$ & $\bullet$ & $\bullet$ & $\bullet$ & $\bullet$ & $\bullet$	\\ 
 \hline
 O$8$-plane 	& --- & --- & $\bullet$ & --- & --- & --- & --- & --- & --- & ---				\\
 \hline
\end{tabular}
\caption{The brane intersection. In the above notation --- denotes that the D-brane or the O-plane extends along that particular direction, while $\bullet$ means that the coordinate is transverse to their worldvolume.}
\label{table: D2_O8}
\end{center}
\end{table}

It has been shown that the supergravity background is described by the invariant line element \cite{Dibitetto:2018ftj}
\begin{equation} \label{eq: exceptional_sugra_1}
ds^2 = \frac{4}{9} \left( \frac{q}{p} \right)^{1/3} \left( \frac{1+r^3}{\sqrt{r}}~ds^2_{AdS_3}+\frac{9}{4}\frac{\sqrt{r}}{1+r^3}~dr^2 + \frac{9}{4} \frac{1}{\sqrt{r}} ds^2_{S^6}  \right)\, ,
\end{equation}
where in the above the six-dimensional unit-sphere can be written explicitly as
\begin{equation}
\begin{aligned}
ds^2_{S^6} &= d \theta^2_1 + \sin^2 \theta_1 (d \theta^2_2 + \sin^2 \theta_2 d \theta^2_3) \\ 
&+ \sin^2 \theta_1 \sin^2 \theta_2 \sin^2 \theta_3 \left(\vphantom{\frac{1}{2}}  d \phi^2_1 + \sin^2 \phi_1 (d \phi^2_2 + \sin^2 \phi_2 d \phi^2_3) \right)\, ,
\end{aligned}
\end{equation}
with the dialton and the field in the R-R sector being equal to:
\begin{equation} \label{eq: exceptional_sugra_main_1}
\begin{aligned}
e^{\phi} 	&= q^{-1/6}~p^{-5/6}~r^{-5/4}\, , \\
F_6 		&= 5~q~\vol_{S^6}\, ,
\end{aligned}
\end{equation}
and the Romans mass is equal to 
\begin{equation} \label{eq: exceptional_sugra_main_2}
F_0 = p\, .
\end{equation}

The analysis of the local supersymmetry constraints leads to the vanishing of the fields in the NS-NS sector.

There are two particularly interesting limits that one can consider. The first is the $r \rightarrow 0$ and the second is the $r \rightarrow \infty$. 

Upon considering the former, the metric and the dilaton become 
\begin{equation}
\begin{aligned}
ds^2		&= \frac{4}{9}~\frac{1}{\sqrt{r}}~ds^2_{AdS_3} + \sqrt{r}~dr^2 + \frac{1}{\sqrt{r}}~ds^2_{S^6}\, , \\
e^{\phi}	&= r^{-5/4}\, .
\end{aligned}
\end{equation}

The above is the metric description of the O$8$ orientifold plane with a diverging dilaton field. This behaviour has been observed in other supergravity descriptions with an $AdS_5$, $AdS_6$ and $AdS_7$ factor \cite{Brandhuber:1999np,Bah:2017wxp}. In spite of the existence of a diverging string coupling, the holographic computations were performed without any obtrusive issues.

The second limit appears more interesting as it presents a special peculiarity. Let us consider the metric and the dilaton of \cref{eq: exceptional_sugra_1} and take the limit $r \rightarrow \infty$. We obtain 
\begin{equation}
\begin{aligned}
ds^2		&= r^{5/2}~ds^2_{AdS_3} + \frac{9}{4}~\frac{1}{r^{5/2}}~\left( dr^2 + r^2~ds^2_{S^6} \right) \, ,\\
e^{\phi}	&= r^{-5/4} \, .
\end{aligned}
\end{equation}

The above resembles a lot the metric of a D$2$-brane when we consider the near-horizon limit \cite{Johnson:2000ch} and with the identification of $r$ as being the radial dimension of the $\mathbb{R}^7$ space that spans the $\{3,4,\cdots,9 \}$ subspace of the ten-dimensional geometry. This is precisely the quaint feature that we mentioned. One would have expected to obtain this limit when taking small values of the $r$ as arises in the expansion of the metric of the standard D$2$-brane and not the other way around as was obtained here. This is perhaps counterintuitive, however it is related to the fact that the metric is not compact. This is also evident from the computation of the central charge in the dual two-dimensional superconformal field theory, a computation that also suggests that the dual SCFT contains infinite degrees of freedom. 

In order to remedy this malady, one can consider gluing two mirrored solutions together. On the locus where the gluing takes place there exists a defect that is interpreted as a D$8$-brane. As was the case in the massive IIA example above, when one crosses the point where the gluing took place the Romans' mass, $F_0$, flips sign (jumps), while the metric and the dilaton field remain continuous.  

After performing the above procedure, the computation of the central charge yields a finite result. 

In \cref{eq: exceptional_sugra_1,eq: exceptional_sugra_main_1,eq: exceptional_sugra_main_2} there are the quantities $p$ and $q$. These are constants and more specifically we have that: 
\begin{equation}
p \in \mathbb{Z}, \qquad \frac{q}{6 \pi^2} \in \mathbb{Z} \, ,
\end{equation}
which are results that follow from the flux quantization.

\begin{figure}[H]
	\begin{center}
	\includegraphics[width=9cm]{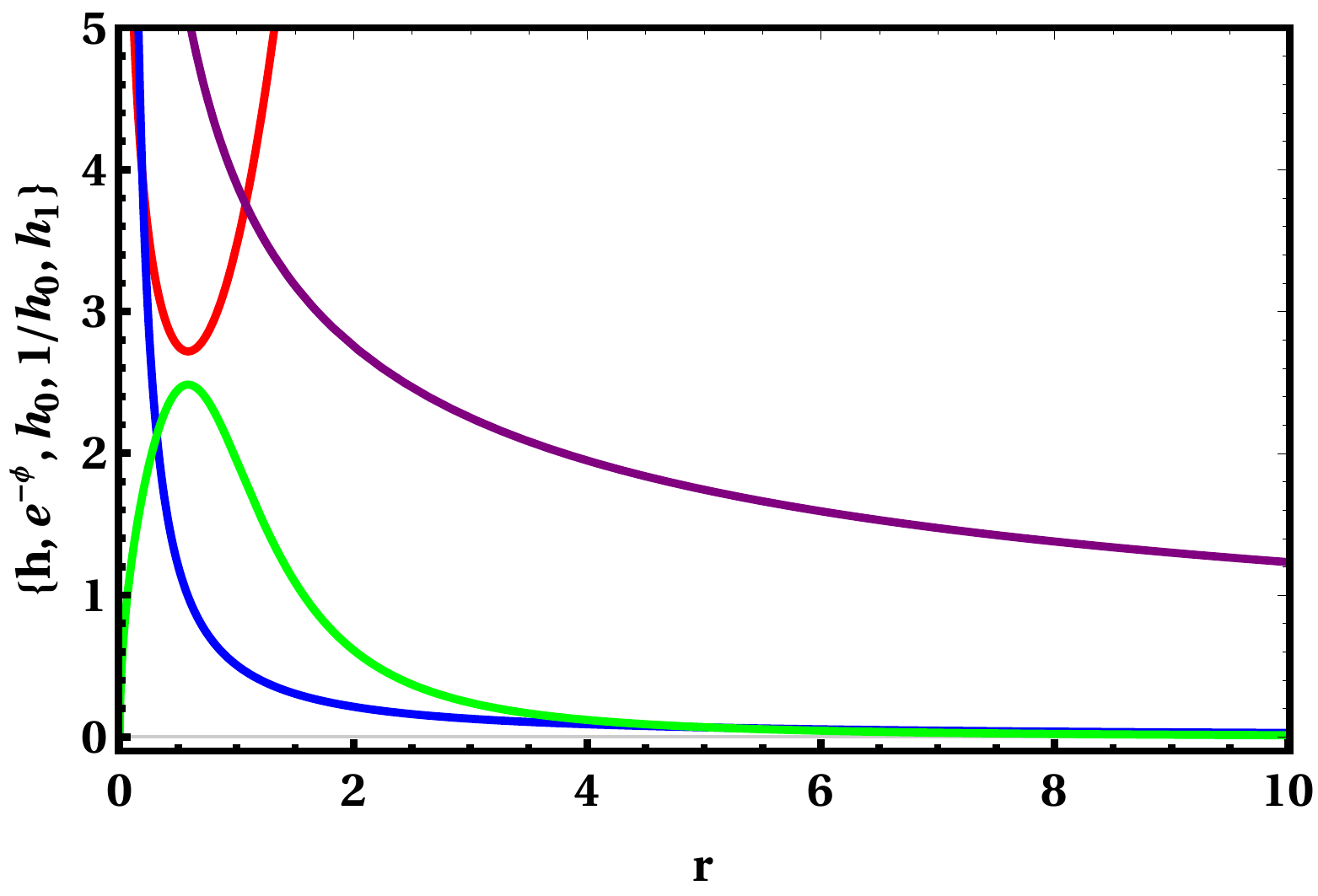} 
	\caption{We show the various warp factors of the metric \cref{eq: exceptional_sugra_1} in the massive IIA theory that realizes the maximal $F(4)$ exceptional supersymmetry. We chose to set $p=1$ and $q=6 \pi^2$. The red line is the $h_0$ factor, the blue line is the $e^{\phi}$, the green line is the $1/h_0$ factor and finally the purple line is the $h_1$ using the labelling of \cref{eq: general_geometry_1}. Observe that $h_0$ and $1/h_0$ differ only by a constant factor. We have accounted for that particular constant factor in the plot.}
	\label{exceptionalIIAF4}
  	\end{center}
\end{figure}

In order to get a better feel on the above supegravity solution, we plot the various warp factors of the metric \cref{eq: exceptional_sugra_1} in \cref{exceptionalIIAF4}

While we have provided the main features of the local solution that interests us in this section, we have been far from exhaustive and explicit. As we have done in previous sections, we urge the interested reader to find and study more details and explicit computations in \cite{Dibitetto:2018ftj} where this solution was orginally obtained. 

We will be working, once again, with the local solution described here as the piecewise character that follows from the gluing of the two mirrored solutions can always be assumed. 

We are interested in the NVEs that follow from the equations of motion. As we have provided a thorough analysis in terms general functions for the warp factors and performed an example explicitly while explaining the classical analogy of the systems, here we work more quickly provide directly the fluctuations of the $\rho$ and $\theta_1$-coordinates.  

We begin by the $\theta_1$ fluctuations while keeping all the other dimensions frozen; their values are fixed on the invariant plane of solutions given by \cref{eq: gen_invariant_plane_1} and evaluated on the simple solutions given by \cref{eq: general_simple_sltn_r_1}. We expand as $\theta_1 = 0 + \varepsilon~\vartheta$ in the infinitesimal limit where $\varepsilon \rightarrow 0$ and work to linear order to obtain 
\begin{equation} \label{eq: NVE_1 _F4}
\ddot{\vartheta} - \frac{1}{2 \tau} ~ \dot{\vartheta} + \alpha^2_2 ~ \vartheta = 0 \, ,
\end{equation}
which can be solved by Bessel functions of the first and the second kind 
\begin{equation}
\vartheta = \tau^{3/4} \left(\vphantom{\frac{1}{2}} c_1~J_{3/4}(\alpha_2 \tau) + c_2~Y_{3/4}(\alpha_2 \tau)  \right)\, ,
\end{equation}
where $c_{1,2}$ are just the two constants of integration.

The Bessel function is Liouvillian for half-integer rank as it can be re-expressed equivalently in terms of simple trigonometric functions and polynomials. Having said that, our chance of spotting a non-Liouvillian solution as a result of the Kovacic algorithm or no solution at all lies in the NVE for the $\rho$-dimension. In order to derive and study the NVE relevant for that dimension we expand as $\rho = 0 + \varepsilon~\varrho$ in the limit $\varepsilon \rightarrow 0$. This yields 
\begin{equation} \label{eq: NVE_2 _F4}
\ddot{\varrho} + \frac{5 E^3 \tau^3-1}{2 \tau(E^3 \tau^3+1)} ~ \dot{\varrho} + \frac{E^3 \tau + (\alpha_1 + \alpha_1 E^3 \tau^3)^2}{(1 + E^3 \tau^3)^2}~\varrho = 0\, .
\end{equation}

The above equation does not admit analytical solutions, hence we deduce that the NVE has a non-Liouvillian solution. Since the NVE for the $\rho$-coordinate demonstrates a non-integrable behaviour we declare the supergrvity description given by \cref{eq: exceptional_sugra_1} as being non-integrable on general grounds\footnote{note that the NVEs here are not parametric differential equations. The only parameter is the wrapping of the string around the spherical coordinate for which we have two physically distinct cases; either $\alpha_1=0$ or $\alpha_1 \neq 0$. Hence the failure to find an analytic solution in terms of quadratures is enough and we do not need to enforce the full power of the Kovacic algorithm.}. 
	\subsection{Comments on the O$8$-plane and D$2$-brane limit} 
As we have already seen, the above maximal local supergravity description possesses two very interesting limits. If we consider $r \rightarrow 0$ then we recover the orientifold O$8$-plane with a diverging dilaton, while in the $r \rightarrow \infty$ we end up with a metric that has all the hallmarks of the standard D$2$-metric. The local solution described by \cref{eq: exceptional_sugra_1} can be thought of as an interpolation between these two limits. 

\begin{figure}[b]
	\begin{center}
	\includegraphics[width=13cm]{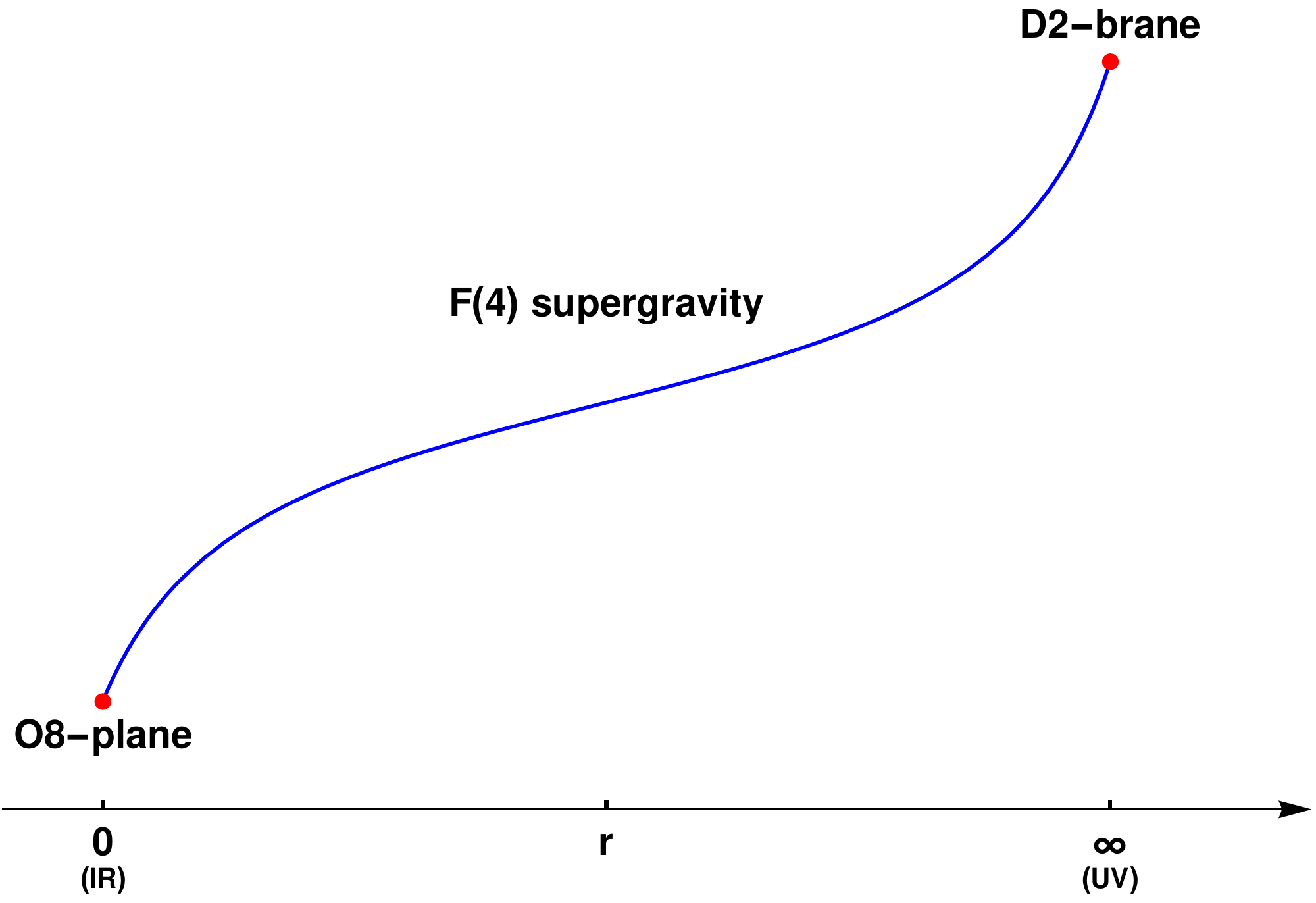} 
	\caption{A cartoon represenation of the $F(4)$ solution as it interpolates from the O$8$-plane in the IR to the D$2$-brane in the UV.}
	\label{IIAF4flow}
  	\end{center}
\end{figure}

Having shown that the exceptional $F(4)$ supergravity solution is non-integrable, an interesting question arises. Is it integrable at its end-points or not? Simply put, is the local solution a non-integrable vacuum that interpolates or perhaps better phrased flows between two integrable ones or not? We show this pictorial representation in \Cref{IIAF4flow}. 

Let us begin by considering the situation in the limit where we obtain the orientifold O$8$-plane. For convenience we reproduce here the invariant line element, which is  
\begin{equation}
\begin{aligned}
ds^2		&= \frac{4}{9}~\frac{1}{\sqrt{r}}~ds^2_{AdS_3} + \sqrt{r}~dr^2 + \frac{1}{\sqrt{r}}~ds^2_{S^6}\, , \\
e^{\phi}	&= r^{-5/4}\, .
\end{aligned}
\end{equation}

The above metric description is in such a form that the general formulae of \cref{sec: no_nsns_sector} are directly applicable. Since we have streamlined the approach of spotting a non-integrable sector in the systems enough times, here we proceed directly to the examination of the NVEs that are associated with the $\theta_1$ and the $\rho$-dimensions. 

The NVE for the $\theta_1$-dimension in the above background becomes, 
\begin{equation}
\ddot{\vartheta} - \frac{1}{2 \tau}~\dot{\vartheta} + \frac{1}{2}~\alpha^2_2~\vartheta = 0\, ,
\end{equation}
which admits as solution a linear combination of Bessel functions of the first and the second-kind. In both cases, the rank of the Bessel function is half-integer and as we have already discussed this means that the solution is Liouvillian. 

The chance of detecting a non-integrable sector falls again on the NVE for the $\rho$ coordinate. For the orientiold plane that we consider here, it reads 
\begin{equation} \label{eq: NVE_rho_O8}
\ddot{\varrho} - \frac{1}{2~\tau}~\dot{\varrho} + \left(\frac{2}{3}~E^3~\tau + \alpha^2_1 \right)~\varrho = 0\, ,
\end{equation}
which does not admit a simple analytic solution in terms of quadratures and hence we have spotted a signature of non-integrability. Thus, we can declare the full O$8$-plane solution as being non-integrable. 

Let us point at this point that the NVE described by \cref{eq: NVE_rho_O8} admits simple solutions in terms of $\sin$ and $\cos$ in the limiting case $\alpha_1 \rightarrow 0$; that is the limit where the string degenerates to a point-particle. 

Finally, we are left to examine whether or not the UV limit of the $F(4)$ supergravity is Liouville integrable; namely the D$2$-brane solution. The authors in \cite{Stepanchuk:2012xi} have already argued that the D$2$-brane solution exhibits non-integrable dynamics, and more generally they showed that this is indeed the case for any intersecting brane background interpolating between the flat space limit and an $AdS_p \times S^p$ spaccetime. 

Finally, the picture of the $F(4)$ supergravity solution is now quite clear. It interpolates, as is shown in \Cref{IIAF4flow}, from a non-integrable vacuum in the IR (the orientifold O$8$-plane) to a non-integrable theory in the UV (the D$2$ solution), while itself exhibits non-integrable dynamics for an extended string. 
 	\subsection{Comments on the low-energy description of the modes}
As we did for the classses of supergravity solutions in the preceding sections, here we are discussing the poin-like  limit of the string embedding in which the wrapping around the cyclic cooridnates is set to zero. We start by taking the two NVEs derived above, \cref{eq: NVE_1 _F4,eq: NVE_2 _F4}, and setting directly $\alpha_1=\alpha_2=0$.

From the former we obtain 
\begin{equation}
\ddot{\vartheta} - \frac{1}{2 \tau} ~ \dot{\vartheta} = 0 \, ,
\end{equation}
which admits the Liouville integrable solution
\begin{equation}
\vartheta = e_1 + e_2~\frac{2}{3}~\tau^{3/2}\, .
\end{equation}

From the latter, \cref{eq: NVE_2 _F4}, we arrive at 
\begin{equation}
\ddot{\varrho} + \frac{5~E^3~\tau^3-1}{2~\tau(E^3~\tau^3+1)}~\dot{\varrho} + \frac{E^3~\tau}{(E^3~\tau^3+1)^2}~\varrho = 0 \, ,
\end{equation}
which also admits a Liouvillian solution given by
\begin{equation}
\varrho = \ell_1~\cos \left(\frac{2}{3}~\ArcTan(E^{3/2}~\tau^{3/2}) \right) + \ell_2~\sin \left(\frac{2}{3}~\ArcTan(E^{3/2}~\tau^{3/2}) \right)\, .
\end{equation}

As in the other backgrounds we examined, it appears that the NVEs for the point particles are Liouville integrable, while the equations corresponding the extended wrapped strings are not. 
	\subsection{Comments on instantonic string configurations}	
As we did in the previous sections, here we also examine the instantonic string modes of zero-energy. It is quite clear that the NVE for the $\theta_1$-dimension \cref{eq: NVE_1 _F4} will not be affected in any way and the said NVE has Liouvillian solutions, since the Bessel functions are of half-integer rank. We examine the $\rho$-NVE, \cref{eq: NVE_2 _F4}, after setting $E=0$ which becomes 
\begin{equation}
\ddot{\varrho} - \frac{1}{2~\tau} ~ \dot{\varrho} + \alpha^2_1 ~ \varrho = 0\, ,
\end{equation}  
which is the same as the $\theta_1$ NVE and hence has Liouvilian solutions. The discussion for the implications of this result is exactly the same as in \cref{sec: insta_IIA}. 
\section{Peroration} \label{sec: final}
The evident outcome of the present work is the definite characterization of certain classes of vacua in the massive IIA and IIB theories as well as the local unique supergravity description with the exceptional $F(4)$ superalgebra as being non-integrable. The class of supergravity descriptions that arose in the context of the massive IIA theory realizes the large $\mathcal{N}=(4,0)$ superconformal symmetry, while the class of solutions in the IIB theory realizes small $\mathcal{N}=(4,0)$. The exceptional $AdS_3$ solution with the $F(4)$ superalgebra is $\mathcal{N}=8$. 

Since we dealt with two classes of solutions that are parameterized in terms of some constants and we ended up having parameterized differential equations, we contemplated the possibility of a potential fine tuning that might provide a classically integrable solution. We were able to conclusively determine that this never occurs. In order to do so, we enforced the analytic algorithm by Kovacic and we examined each and every step of the said algorithmic construction. Having been able to spot a non-integrable sector in the unique $F(4)$ local solution we can safely declare that this particular supergravity vacuum, and also the associated global solution after gluing two mirrored copies together, is non-integrable. 

Our findings are summarized in the following

\begin{itemize}
\item \textbf{Conclusion A}: Every supergravity solution derived from the massive IIA solution with the large $\mathcal{N}=(4,0)$ supersymmetry which is described in \cref{sec: largesusy} is non-integrable. 
\item \textbf{Conclusion B}: Every supergravity solution derived from the IIB solution with the small $\mathcal{N}=(4,0)$ which is described in \cref{sec: IIB_vacua} is non-integrable. 
\item \textbf{Conclusion C}: The local $AdS_3$ solution with the $\mathcal{N}=8$, exceptional $F(4)$ symmetry described in \cref{sec: exceptional} is non-integrable. 
\end{itemize}

In the two families of the supergravity solutions that we examined there are two limiting cases where the metric descriptions experience a supersymmetry enhancement to $\mathcal{N}=(4,4)$. In these two limiting cases the local supergravity descriptions return to two theories which are known to be integrable. The massive IIA class returns to the familiar $AdS_3 \times S^3 \times \tilde{S}^3 \times S^1$ and the IIB class becomes the $AdS_3 \times S^3 \times T^4$ solution.

In each section we also considered the point-like limit of the string. This is setting in which the wrapping of the string around the cyclic coordinates is set to zero. In addition to that, we also examined instantonic string configurations of zero energy and we discussed the implications of our results. 

All of the above supergravity descriptions have a common feature, viz. the appearance of non-trivial warp factors. In this work we have seen that the integrable cases arise when the warping factors become mere constants rather than non-trivial functions. This statement does not constitute a formal proof of course. It serves as supporting evidence to the prevalent token. This phenomenon has appeared a number of times in the existing literature. The Sfetsos-Thompson vacuum \cite{Nunez:2018qcj,Sfetsos:2010uq} is the unwarped solution in the core of the Gaiotto-Maldacena $AdS_5$ class of supergravity descriptions. Within the discovered family of the massive IIA $AdS_7$ vacua \cite{Apruzzi:2013yva,Apruzzi:2015wna} there exists the recent construction in \cite{Filippas:2019puw}. More evidence for the relation between integrability and the unwarping is also provided in \cite{Filippas:2019ihy} for the class of supergravity solutions obtained in \cite{Lozano:2019emq,Lozano:2019zvg}.
\section*{A\lowercase{cknowledgements}}
I am delighted to thank Kostas Filippas for insightful discussions and recommendations, as well as Nial Macpherson and Achilleas Passias for useful correspondence. It is also a pleasure to thank Dimitris Giataganas for enlightening comments and suggestions in addition to a crucial read and remarks on the final draft of this work.  
\setcounter{equation}{0}
\appendix
\newpage
\section{NVEs and the analytic Kovacic algorithm} \label{app: analytic_kovacic} 
As we have seen, all the NVEs can be brought in the general form 
\begin{equation} \label{eq: app_1}
\ddot{x}(\tau) + \mathcal{A}_1(\tau)~\dot{x}(\tau) + \mathcal{A}_2(\tau) ~ x(\tau) = 0\, .
\end{equation}
Under the change of variables 
\begin{equation} \label{eq: change_of_vars}
x(\tau) = e^{\frac{1}{2} \int \mathcal{A}_1(\tau) ~ d\tau}~z(\tau)\, ,
\end{equation}
\cref{eq: app_1} can be brought in the form 
\begin{equation} \label{eq: app_main}
\ddot{z}(\tau) + \mathcal{V}(\tau)~z(\tau) = 0\, .
\end{equation}

In the above equation, the potential $\mathcal{V}$ is given by:
\begin{equation}
\mathcal{V} = \mathcal{A}_2 - \frac{1}{2} ~ \dot{\mathcal{A}_1} - \frac{1}{4} ~ \mathcal{A}^2_1\, .
\end{equation}

After considering the aforementioned change of variables, $x$ is Liouvillian if and only if $z$ is Liouvillian. Thus we are working without loss of generality and additionally, we managed to get rid of the first derivative term of the original equation.

The theorem that Kovacic proved \cite{KOVACIC19863} has three different cases, which are presented below
\begin{itemize}
\item \textbf{Case I}: Every pole of the potential must be of even order or else of order $1$. The order of the potential at infinity must be even or else greater than $2$.
\item \textbf{Case II}: The potential must have at least one pole that is either of odd order greater than $2$ or else of order $2$.
\item \textbf{Case III} The order of a pole of the potential cannot be greater than $2$ and the order of the potential at infinity must be at least $2$.
\end{itemize}

In the case that a differential equation fails to meet all three conditions as presented above, it is enough to declare the differential equation given by \cref{eq: app_main} as being Liouville non-integrable.

On the other hand, if any of the aforementioned conditions is satisfied, then the associated case might hold true. This, in turn, implies that a Liovillian solution exists. Ergo, every time that a condition is satisfied we have to delve into the specific algorithmic steps of the said case. These steps have been formally developed by Kovacic. By examining a differential equation in this way we succeed in doing the following: if a Liouville integrable solution exists, the procedure will give us the answer.

We do not describe the general steps of each case that we encounter here explicitly as this has already been done in \cite{Filippas:2019ihy} in a physics context. The point of this appendix is to show that after enforcing the full power of the analytic Kovacic algorithm by hand, the classes of backgrounds in the massive IIA and IIB theories do not have any special values of the free parameters that give integrable string solutions.  
\numberwithin{equation}{subsection}
\subsection{Aplication in massive IIA with large $\mathcal{N}=(4,0)$ superconformal symmetry} \label{sec: largesusy_app}
We begin the explicit considerations of this appendix by looking at the massive IIA class of backgrounds that realizes large $\mathcal{N}=(4,0)$ superconformal symmetry, \cref{sec: largesusy}. We are dealing with the $\rho$-NVE given by \cref{eq: NVE_rho_massiveIIA_large}. We change variables as shown in \cref{eq: change_of_vars}, and derive an equation
\begin{equation}
\ddot{z} = \mathcal{V}~z\, ,
\end{equation}
with the potential being given by 
\begin{equation} \label{eq: potential_IIA}
\mathcal{V} = \frac{\mathcal{N}_{IIA}(c,F_0,E,\tau^3)}{\left(\tau + \frac{c~q}{L~F_0~E~\nu} \right)^2} \, ,
\end{equation}
where in the above, the numerator is explicitly 
\begin{equation}
\begin{aligned} \label{eq: potential_IIA_num}
\mathcal{N}_{IIA}(c,F_0,E,\tau^3) = \frac{5}{16} - \frac{(c~q+L~F_0~\nu~E~\tau)^2(c~q~E^2+q~\alpha^2_1+L~F_0~\nu~E^3~\tau)}{L^2~q~F^2_0~\nu^2E^2}
\end{aligned}
\end{equation} 

This potential is a candidate only the second case of the Kovacic algorithm. We will work our way through each step of the said case and try to understand whether or not we cna obtain a Liouvillian solution by tuning the parameters appropriately. 

It is evident from \cref{eq: potential_IIA,eq: potential_IIA_num} that the potential has a single pole of order two at 
\begin{equation}
\tau = - \frac{c~q}{L~F_0~\nu~E}\, .
\end{equation}

We perform the expansion as $\tau \rightarrow \infty$ and obtain 
\begin{equation}
\mathcal{V}^{\tau \rightarrow \infty} = - \frac{L~F_0~\nu~E^3}{q}~\tau - (\alpha^2_1+c~E^2) + \frac{5}{16}~\frac{1}{\tau^2} + \mathcal{O}(\tau^{-3}) \, ,
\end{equation}
and it i clear that the order of $\mathcal{V}$ is $-1$\footnote{since the order of the potential is odd, one might be tempted to disregard it at this point and not enforce the step-by-step procedure. However, we should be careful as we do not have a simple polynomial of odd degree, which indeed according to Kovacic has no Liouvillian solution, but rather a fraction of polynomials. Hence, since the potential is a good candidate for a certain case, we should follow the steps of that case explicitly. We are grateful to Kostas Filippas for clarifying this subtle point and encouraging us to do the computation.}. 
	\subsubsection{Case II}
We continue the analysis here by re-expressing the potential \cref{eq: potential_IIA} in terms of partial fractions. This yields 
\begin{equation} \label{eq: partial_frac_IIA}
\mathcal{V} = - (\alpha^2_1 + c~E^2) - \frac{L~F_0~\nu~E^3}{q}~\tau+ \frac{5/16}{\left(\tau + \frac{c~q}{L~F_0~\nu~E} \right)^2}\, .
\end{equation}

We proceed to the construction of the new coordinates which are given by 
\begin{equation}
\mathcal{E}_i = \{2 + k ~ \sqrt{1 + 4 ~ \beta_i}|k=0,\pm 2 \} \cap \mathbb{Z}\, ,
\end{equation}
where in the above $\beta_i$ is the coefficient of the pole term $1/(\tau-\tau_i)$. In this case, we obtain 
\begin{equation}
\mathcal{E}_0 = \left\lbrace ~ 2 - 2 \sqrt{1+4 \frac{5}{16}}, ~ 2 ~, 2 + 2 \sqrt{1+4 \frac{5}{16}}  \right\rbrace = \{-1,~2,~5 \}\, .
\end{equation}

Since the order of the potential at infinity is minus one, we also have the number $\mathcal{E}_{\infty}=\{-1\}$. Now, we define the quantity $d$ to be given via 
\begin{equation}
d = \frac{1}{2} ~ \left(\vphantom{\frac{1}{2}} e_{\infty}-\sum_i e_i \right)\, ,
\end{equation}
where in the above we have $(e_c)_{c \in \Gamma \cup (\infty)}$ with $e_c \in E_c$ and $\Gamma$ is the set of poles of the potential under examination.

For the case that we consider here, $e_{\infty}=-1$ always holds. Hence, we are left with only one free numbers to take values from the set of $\mathcal{E}$-coordinates that we computed above. There are $3$ possible combinations for which we evaluate explicitly the quantity $d$. All the results are shown below 
\begin{equation}
d = 
	\begin{cases}
	0\, , \\
	-3/2\, ,\\
	-3\, .
	\end{cases}
\end{equation}

According to the theorem by Kovacic, in order for us to be able to proceed in our quest for Liouville integrable solutions, the above result should be a non-negative integer  number, $d \in \mathbb{Z}_{\geq 0}$. If $d$ fails to satisfy this condition, then the algorithm stops here and no Liouvillian solutions exist. From all of the above results, only $d=0$ is a viable result, otherwise the algorithm stops here. 

We now move on to the next step and we construct the rational function 
\begin{equation}
\mathcal{X} = \frac{1}{2} ~ \sum_i \frac{e_i}{\tau - \tau_i}\, ,
\end{equation}
where in the above we consider the particular values of the various $e_i$ that gave the result $d=0$; namely $e_0=-1$. Hence, the rational function defined above for our example is equal 
\begin{equation} \label{eq: rational_fnctn_IIA}
\mathcal{X} = - \frac{1}{2} ~ \frac{1}{\left(\tau + \frac{c~q}{L~F_0~\nu~E} \right)^2} \, .
\end{equation}

Having obtained the above, we now seek for a monic polynomial of degree $0$ (the value of $d$) that satisfies the equation
\begin{equation}
\dddot{\mathcal{P}} + 3~\mathcal{X}~\ddot{\mathcal{P}} + (3~\mathcal{X}^2+3~\dot{\mathcal{X}}-4~\mathcal{V})~\dot{\mathcal{P}} + (\ddot{\mathcal{X}}+3~\mathcal{X}~\dot{\mathcal{X}} + \mathcal{X}^3 - 4~\mathcal{V}~\mathcal{X} - 2~\dot{\mathcal{V}})~\mathcal{P} = 0\, .
\end{equation} 

However, in the above we have $\mathcal{P}=1$ and therefore the statement reduces to 
\begin{equation}
\ddot{\mathcal{X}}+3~\mathcal{X}~\dot{\mathcal{X}} + \mathcal{X}^3 - 4~\mathcal{V}~\mathcal{X} - 2~\dot{\mathcal{V}} = 0\, .
\end{equation}

We insert, in the above condition, the result given in \cref{eq: rational_fnctn_IIA} for $\mathcal{X}$ as well as \cref{eq: partial_frac_IIA} for the potential $\mathcal{V}$. The result of this is of substancial size and we do not write  it explicitly, since it is quite straightforward to derive it using any mathematical software. The interesting part is what the result implies. 

A close inspection reveals that in order to satisfy the equation there are the following choices
\begin{equation}
E=0, \quad \textsf{or}, \quad F_0 = 0, \quad \textsf{or}, \quad L=0, \quad  \textsf{or} \quad \nu=0\, .
\end{equation}

We have already discussed the $E=0$ which is the instantonic string mode in \cref{sec: insta_IIA} and what this solution implies. This is not a suggestion that the string sector is integrable, as integrability should be a universal property and for $E \neq 0$ this does not occur. Likewise, we are interested in the massive IIA and the Romans' mass should not be vanishing. The $F_0=0$ is a very special limit, which we discussed in \cref{sec: IIA_enhancement} and is known to be integrable, however it resides outside the massive IIA realm. Finally, the other two solutions are ruled out by the string theory setup of \cref{sec: largesusy}.
\subsection{Aplication in IIB with small $\mathcal{N}=(4,0)$ superconformal symmetry} \label{sec: smallsusy_app}
Let us consider the IIB class of backgrounds that preserves small $\mathcal{N}=(4,0)$ superconformal symmetry \cref{sec: IIB_vacua} and more specificaly the NVE for the $\rho$-dimenson given by \cref{eq: NVE2_IIB}. Performing the analysis as shown in \cref{app: analytic_kovacic}, we can bring the NVE in the form \cref{eq: app_main} and we derive the relevant potential
\begin{equation} \label{eq: potential_IIB}
\mathcal{V} = \frac{\mathcal{N}_{IIB}(c_1,c_2,E,\tau^6)}{\tau^2 \left(\tau - \frac{i}{c_2~E} \sqrt{\frac{c_1}{a}} \right)^2\left(\tau + \frac{i}{c_2~E} \sqrt{\frac{c_1}{a}} \right)^2} \, ,
\end{equation}
with the numerator being explicitly given by 
\begin{equation} \label{eq: potential_IIB_num}
\begin{aligned}
\mathcal{N}_{IIB}(c_1,c_2,E,\tau^6) = &-(\alpha^2_1 + a~c_2~E^2)~\tau^6 - 3~\frac{c_1}{c_2}~\tau^4 - \frac{c^2_1(1+4~\alpha^2_1~\tau^2)}{4~a^2~c^2_2~E^4} \\ 
&-\frac{c^2_1(c_1+3~a~c^2_2~E^2~\tau^2)}{a^2~c^5_2~E^4} - \frac{c_1(3+4~\alpha^2_1~\tau^2)}{4~a^2~c^2_2~E^4}
\, .
\end{aligned}
\end{equation}

It is obvious that the potential derived above satisfies the first and second cases and below we are going through the steps of each one of those cases explicitly. 

The poles of the potential \cref{eq: potential_IIB} are at 
\begin{equation} \label{eq: poles_IIB}
\tau = 0\, ,\quad \tau = \pm \frac{i}{c_2~E} \sqrt{\frac{c_1}{a}} \, ,
\end{equation}
and are of order two. The expansion of the potential around infinity is 
\begin{equation} \label{eq: infinity_IIB}
\mathcal{V}^{\tau \rightarrow \infty} = -(\alpha^2_1 + a~c_2~E^2) - \frac{c_1}{c_2} ~ \frac{1}{\tau^2} - \frac{3~c_1}{2~a~c^2_2~E^2} ~ \frac{1}{\tau^4}+ \mathcal{O}(\tau^{-6}) \, ,
\end{equation}
exhibiting a zeroth order behaviour there. 
	\subsubsection{Case I}
We continue the analysis of this section, by re-expressing the potential \cref{eq: potential_IIB} in terms of partial fractions. This yields 
\begin{equation} \label{eq: partial_frac_IIB}
\begin{aligned}
\mathcal{V} = &-(\alpha^2_1 + a~c_2~E^2) - \left(\frac{1}{4} + \frac{c_1}{c_2} \right) \frac{1}{\tau^2} + \frac{5/16}{\left(\tau + \frac{i}{c_2~E} \sqrt{\frac{c_1}{a}} \right)^2}+ \frac{5/16}{\left(\tau - \frac{i}{c_2~E} \sqrt{\frac{c_1}{a}} \right)^2} \\
&- \frac{3 ~ i ~ c_2 ~ E}{16}~\sqrt{\frac{a}{c_1}}~\frac{1}{\tau + \frac{i}{c_2~E}\sqrt{\frac{c_1}{a}}} +\frac{3 ~ i ~ c_2 ~ E}{16}~\sqrt{\frac{a}{c_1}}~\frac{1}{\tau - \frac{i}{c_2~E}\sqrt{\frac{c_1}{a}}} \, .
\end{aligned}
\end{equation} 

We want to construct the complex numbers given by 
\begin{equation}
a^{\pm}_{i} = \frac{1}{2} \pm \frac{1}{2} \sqrt{1 + \beta_i} \, ,
\end{equation}
where the $\beta_i$ are the coefficients of the $1/(\tau-\tau_i)^2$ pole terms in the partial fractions expansion, see \cref{eq: partial_frac_IIB}. For brevity and notational convenience we shall call $i=0$ the $\tau=0$ pole and use $i=\{1,2\}$ for the $+$ and the $-$ poles in \cref{eq: poles_IIB} respectively. We obtain 
\begin{equation}
\begin{aligned}
a^{\pm}_0 &= \frac{1}{2} \pm \frac{1}{2} \sqrt{1 - 4 \frac{c_1}{c_2}}\, , \\
a^{\pm}_1 = a^{\pm}_2 &= 
						\begin{cases} 
      						&5/4\, , \\
      						&-1/4  
   						\end{cases}
\end{aligned}
\end{equation}

We proceed by examining the data related to form of the potential at infinity. We define a rational function, which we denote by $[\sqrt{\mathcal{V}}]^{\infty}$ , which should be just a number since the order of the asymtotic expansion at infinity is zero, see \cref{eq: infinity_IIB}. Hence, if we write $[\sqrt{\mathcal{V}}]^{\infty} = \breve{a}$ and we perform the matching to the expression \cref{eq: infinity_IIB} we obtain 
\begin{equation}
\breve{a} = i ~ \sqrt{\alpha^2_1 + a ~ c_2 ~ E^2}\, .
\end{equation}
Using the above we can construct the numbers
\begin{equation}
a^{\pm}_{\infty} = \pm \frac{\beta_{\infty}}{2 \breve{a}}\, ,
\end{equation}
where $\beta_{\infty}$ is the coefficient of the $1/\tau$ term in the asymptotic expansion around infinity and hence in our case is zero. 

Now, we proceed and we consider the numbers given by 
\begin{equation} \label{eq: def_d_IIB}
d = a^{\sign(\infty)}_{\infty} - \sum_{i} a^{\sign(i)}_i\, ,
\end{equation}
where in the above we use the symbol $\sign(\#)$ to denote the different possible signs. 

We account for all possible sign contributions which are $2^3$ and finally our results can be summarized as follows:
\begin{align}
d = 
			\begin{cases}
			&3 \pm \frac{1}{2} ~ i ~ \sqrt{4 \frac{c_1}{c_2}-1}\, , \\
			&\frac{3}{2} \pm \frac{1}{2} ~ i ~ \sqrt{4 \frac{c_1}{c_2}-1}\, , \\
			&\hphantom{\frac{3}{2}} \pm \frac{1}{2} ~ i ~ \sqrt{4 \frac{c_1}{c_2}-1}\, .
			\end{cases}
\end{align}

According to the theorem by Kovacic, in order to find integrable solutions the above result should be a non-negative integer  number, $d \in \mathbb{Z}_{\geq 0}$. If $d$ fails to satisfy this condition, then the algorithm stops here and there are no Liouvillian solutions. Thus, the reality of $d$ is a necessary, though not sufficient, condition for integrability. 

The above discussion leaves only two alternatives. Either $d=3$ or $d=0$, both of which amount to choosing $c_2=4~c_1$. 

We consider the possibility $d=0$ first. 

Following the next step that we need to perform as dictated by Kovacic, a general candidate for $\Omega$ is given by the general formula
\begin{equation}
\Omega = \sum_{p}\left( [\sqrt{\mathcal{V}}]^p + \frac{a^{\sign(p)}_p}{\tau-p} \right) + \sign(\infty) [\sqrt{\mathcal{V}}]^{\infty}\, ,
\end{equation} 
where $p$ denotes a certain pole and we are summing over all poles. 

However, in our case the $\tau=0$ is a pole of order two which sets $[\sqrt{\mathcal{V}}]^0 = 0$ and the same holds for the other poles as well. Consequently, we have 
\begin{equation}
\Omega = \frac{1}{2 \tau} - \frac{1}{4 \left(\tau + \frac{i}{4c_1~E} \sqrt{\frac{c_1}{a}} \right)} - \frac{1}{4 \left(\tau - \frac{i}{4c_1~E} \sqrt{\frac{c_1}{a}} \right)} \pm i \sqrt{\alpha^2_1 + 16 ~ a ~ c_1 ~ E^2}\, .
\end{equation} 
Now, again following the steps as dictated by Kovacic, we are looking for a monic polynomial of degree zero and denoted by $\mathcal{P}$, such that 
\begin{equation} \label{eq: monic_poly_general}
\ddot{\mathcal{P}} + 2~\Omega~\dot{\mathcal{P}} + (\dot{\Omega}+\Omega^2-\mathcal{V})~\mathcal{P} = 0\, .
\end{equation}
Since, $\mathcal{P}=1$ the above condition reduces to 
\begin{equation}
\dot{\Omega}+\Omega^2-\mathcal{V} = 0\, ,
\end{equation}
which can never be satisfied as the coefficient in front of the $1/\tau^2$ term is equal to $1/4$.

Now we move on to examine the two families that result in $d=3$. These are the $\{a^{\pm}_0, a^{+}_1, a^{+}_2\}$. For this case, a candidate for $\Omega$ is given by 
\begin{equation}
\Omega = \frac{1}{2 \tau} + \frac{5}{4 \left(\tau + \frac{i}{4c_1~E} \sqrt{\frac{c_1}{a}} \right)} + \frac{5}{4 \left(\tau - \frac{i}{4c_1~E} \sqrt{\frac{c_1}{a}} \right)} \pm i \sqrt{\alpha^2_1 + 16 ~ a ~ c_1 ~ E^2}\, .
\end{equation} 

We are now looking for a cubic monic polynomial $\mathcal{P}=\tau^3 + w_2~\tau^2+w_1 \tau + w_0$, and check if there are values that solve \cref{eq: monic_poly_general}. While the computation is a bit more involved than the previous one, after inserting the expressions appropriately in \cref{eq: monic_poly_general} we deduce that it can never be satisfied and thus these families also fail to provide Liouvillian solutions. 
	\subsubsection{Case II}
Since the potential given in \cref{eq: potential_IIB,eq: potential_IIB_num} also satisfies the criteria for the second case and we already excluded Case I, we now shift our focus to the algorithmic steps of finding an analytic solution in terms of quadratures in case II. We have done much of the work already in the previous section, namely deriving the potential, expressing it in partial fractions and expanding it near infinity. 

We have also computed the coefficients $\beta_i$ that are derived from looking at the pole terms. Now, in close relation to the previous procedure, we want to build new coordinates using the coefficients of the pole terms, $\beta_i$, which are given by 
\begin{equation}
\mathcal{E}_i = \{2 + k ~ \sqrt{1 + 4 ~ \beta_i}|k=0,\pm 2 \} \cap \mathbb{Z}\, .
\end{equation}

We can now use our values for $\beta_{0,1,2}$ and obtain 
\begin{equation*}
\mathcal{E}_0 = \left\lbrace 2 - 4~i~\sqrt{\frac{c_1}{c_2}},~ 2,~ 2 + 4~i~\sqrt{\frac{c_1}{c_2}}\right\rbrace\, ,\qquad \mathcal{E}_1 = \mathcal{E}_2 = \{-1, 2, 5 \}\, .
\end{equation*}

From the string theory considerations of \cref{sec: IIB_vacua} we know that $c_1 /c_2 >0$ and this excludes the first and third result of $\mathcal{E}_0$. Hence, we are left with the numbers 
\begin{equation}
\mathcal{E}_0 = \{ 2\}, \qquad \mathcal{E}_1 = \mathcal{E}_2 = \{-1, 2, 5 \}\, .
\end{equation}

Since the order of the potential at infinity is zero we also have the number $\mathcal{E}_{\infty}=\{0\}$. In analogy to what we did in the previous case, we define the quantity $d$ to be given via 
\begin{equation}
d = \frac{1}{2} ~ \left(\vphantom{\frac{1}{2}} e_{\infty}-\sum_i e_i \right)\, ,
\end{equation}
where in the above we have $(e_c)_{c \in \Gamma \cup (\infty)}$ with $e_c \in E_c$ and $\Gamma$ is the set of poles of the potential under examination.

For our case $e_{\infty}=0$ always holds as well as $e_0=2$. Hence, we are left with only two free numbers to take values from the set of $\mathcal{E}$-coordinates that we computed above. There are $3^2$ combinations for which we evaluate explicitly the quantity $d$. All the results are grouped below 
\begin{equation}
d = 
	\begin{cases}
	0\, , \\
	-3/2\, ,\\
	-3\, , \\
	-9/2\, ,\\
	-6
	\end{cases}
\end{equation}

According to Kovacic, in order for us to be able to proceed and seek integrable solutions, the above result should be a non-negative integer  number, $d \in \mathbb{Z}_{\geq 0}$. If $d$ fails to satisfy this condition, then the algorithm stops here and there are no Liouvillian solutions. From all of the above results, only $d=0$ is a viable result, otherwise the algorithm stops here. 

We now move on to the next step and we construct the rational function 
\begin{equation}
\mathcal{X} = \frac{1}{2} ~ \sum_i \frac{e_i}{\tau - \tau_i}\, ,
\end{equation}
where in the above we consider the particular values of the various $e_i$ that gave the result $d=0$; namely $e_1=e_2=-1$. Hence, the rational function defined above for our example is equal 
\begin{equation} \label{eq: rational_fnctn_IIB}
\mathcal{X} = \frac{1}{\tau} - \frac{1}{2 \left( \tau - \frac{i}{c_2~E} ~ \sqrt{\frac{c_1}{a}} \right)} - \frac{1}{2 \left( \tau + \frac{i}{c_2~E} ~ \sqrt{\frac{c_1}{a}} \right)}\, .
\end{equation}

Having obtained the above, we now seek for a monic polynomial of degree $0$ (the value of $d$) that satisfies 
\begin{equation}
\dddot{\mathcal{P}} + 3~\mathcal{X}~\ddot{\mathcal{P}} + (3~\mathcal{X}^2+3~\dot{\mathcal{X}}-4~\mathcal{V})~\dot{\mathcal{P}} + (\ddot{\mathcal{X}}+3~\mathcal{X}~\dot{\mathcal{X}} + \mathcal{X}^3 - 4~\mathcal{V}~\mathcal{X} - 2~\dot{\mathcal{V}})~\mathcal{P} = 0\, .
\end{equation} 

However, in the above we have $\mathcal{P}=1$ and therefore the statement reduces to 
\begin{equation}
\ddot{\mathcal{X}}+3~\mathcal{X}~\dot{\mathcal{X}} + \mathcal{X}^3 - 4~\mathcal{V}~\mathcal{X} - 2~\dot{\mathcal{V}} = 0\, .
\end{equation}

We insert, in the above condition, the result given in \cref{eq: rational_fnctn_IIB} for $\mathcal{X}$ as well as \cref{eq: partial_frac_IIB} for the potential $\mathcal{V}$. The result of this is 
\begin{equation}
\frac{4~c_1~\alpha^2_1}{c_1~\tau + a~c^2_2~E^2~\tau^3} = 0\, .
\end{equation}

The above yields the obvious solutions $\alpha_1 = 0$ and/or $c_1 = 0$, none of which is acceptable. The reason is that the first amounts to considering the point-like limit of the string. In other words, for the said choice we do not have an extended wrapped string embedding. The latter is the special limit where the metric returns to $AdS_3 \times S^3 \times T^4$ solution. Both of these have been explicitly studied in \cref{sec: point_like_IIB,sec: IIB_enhancement} respectively. 

We have ultimately proved, after examining each and every step of both potential cases of the analytic Kovacic algorithm, that there is no choice of parameters for which the NVE of $\rho$-coordinate yields an integrable solution. 
\clearpage
\bibliography{lit}{}
\bibliographystyle{utphys}

\end{document}